\documentclass[aps,prl,twocolumn,amsmath,amssymb,footinbib,showpacs,longbibliography,superscriptaddress]{revtex4-2}

\usepackage{times}

\usepackage[english]{babel}
\usepackage{latexsym}
\usepackage{graphics}
\usepackage{subfigure}
\usepackage{epsfig}
\usepackage{color}
\usepackage{hyperref}
\usepackage{braket} 
\usepackage[T1]{fontenc}
\usepackage[latin9]{inputenc}
\usepackage{amstext}
\usepackage{amssymb}
\usepackage{graphicx}
\usepackage{esint}
\usepackage{braket}
\usepackage{babel}
\usepackage{amsmath}
\usepackage{ulem}
\hypersetup{
colorlinks=true,
citecolor=blue,
linkcolor=red,
urlcolor=black
}


\newcommand{\think}[1]{{\color{red} {#1}}}

\newcommand{\ie}{i.e.}

\newcommand{\Er}{E_{\textrm{r}}}

\newcommand{\as}{a_{\textrm{\tiny s}}}

\newcommand{\lettersection}[1]{\paragraph*{#1.---}}

\begin{document}

\title{
Quench spectroscopy for Lieb-Liniger bosons in the presence of harmonic trap
}

\author{Jiachen Yu}
\homepage{These authors contributed equally to this work.}
\affiliation{School of Electronics, Peking University, Beijing 100871, China}

\author{Yuanzhe Hu}
\homepage{These authors contributed equally to this work.}
\affiliation{School of Electronics, Peking University, Beijing 100871, China}

\author{Wenhan Chen}
\affiliation{School of Electronics, Peking University, Beijing 100871, China}

\author{Jianing Yang}
\affiliation{School of Electronics, Peking University, Beijing 100871, China}

\author{Xuzong Chen}
\homepage{xuzongchen@pku.edu.cn}
\affiliation{School of Electronics, Peking University, Beijing 100871, China}
\affiliation{School of Computing and Artificial Inteligence, Fuyao University of Science and Technology}

\author{Hepeng Yao}
\homepage{hepeng.yao@pku.edu.cn}
\affiliation{School of Electronics, Peking University, Beijing 100871, China}

\date{\today}

\think{
\begin{abstract}
Quench spectroscopy has emerged as a novel and powerful technique for probing the energy spectrum of various quantum phases for quantum systems from out-of-equilibrium dynamics. 
While its efficacy has been demonstrated in the homogeneous systems theoretically, most experimental setups feature a confining potential, such as a harmonic trap, which complicates the practical implementations.
In this work, we experimentally probe the quench spectroscopy for one-dimensional bosons in optical lattices with the presence of a harmonic trap, and comparing our results with the density matrix renormalization group simulation. For the Mott insulator phase, although a gap is still observed, the band signal is broadened along the frequency space and cut at the half Brillouin zone, which can be explained by the nearest-neighbor tunneling excitations under harmonic confinement. Comparing with the superfluid spectrum, we can see a clear distinction between the two phases and find the inverse quench with larger amplitude yields the clearest spectrum. 
Our work offers pivotal insights into conducting quench spectroscopy effectively in practical systems.
\end{abstract}
}

\maketitle

\lettersection{Introduction}

Spectroscopy methods are widely used to detect the excitation information of strongly-correlated systems ~\cite{damascelli2004, roux2013, ernst2009, Gritsev-QuenchSpectroscopy-2007, gupta2003, Yao2024}, which reveals physical properties like electronic conductivity, magnetic ordering, superconductivity and superfluidity. The basic idea is to induce an excitation to the system and measure the corresponding response, from which one can extract the demanded spectrum information and dynamical properties of the system. In physical experiments, spectroscopy methods are often realized by different pump-probe techniques, such as angle-resolved-photoemission spectroscopy (ARPES)~\cite{damascelli2004, stewart2008} and Neutron scattering~\cite{Han2012}.


Quantum phase transition in strongly-correlated systems is a crucial topic in quantum simulation. Various detection techniques have been developed to distinct different quantum phases and probe their excitation properties. In cold atom systems, time-of-flight imaging (TOF) is the most widely used one, which measures the momentum distributions and provides the information about quantum coherence. Judging from its interference pattern, one can distinguish the Mott insulator (MI) from the superfluid (SF) phase~\cite{Altman2004, davis1995, richard2003, stoferle-transition-2004, gerbier2005}.
Other techniques such as the band mapping~\cite{greiner2002, spielman2007, jordens2008, Qihuang2023} or the quantum gas microscope~\cite{Cheneau2012, greif2016, Nelson2007, sherson2010, bakr2009, weitenberg2011, haller2015}, allow for further detecting the information about the band and the spatial distributions.
On the spectroscopy side, Raman spectroscopy enables the probing of low-energy excitation spectra~\cite{Devereaux2007,haller-pinning-2010}, while Bragg spectroscopy detects the dynamical structure factor and reveal systems' response spectra to density perturbations~\cite{fabbri2015, Altunta__2025}. In addition, there are novel and powerful approaches such as amplitude modulation spectroscopy which measures excitation spectra~\cite{stoferle-transition-2004,Endres2012}, and measurements of critical velocity using moving optical lattices which determines the onset of dissipation and superfluid breakdown~\cite{JCMun2007}.


In recent years, quench spectroscopy was proposed theoretically~\cite{villa2019, villa2020, PRA2025_JDespres} which appears to be a powerful technique for detecting the excitation spectrum of various quantum phases, including the MI, SF and even Bose glass phase in disordered systems~\cite{giamarchi1987,giamarchi1988, derrico-BG-2014, Zhaoxuan2023,yao-boseglass-2020,PRL2017_YanMi,PRA2017_YanMi,arXiv2025_Phil}. Starting with a global quench, one subsequently measures the post-quench evolution from which one can extract all elementary excitation information. In contrast to standard pump-probe techniques, it employs equal-time commutators to directly extract the dispersion relation of elementary excitations, which is experimentally more accessible. Moreover, it permits comprehensive spectral measurement in a single experimental run, eliminating the need for tedious momentum-space scanning.  However, current research on quench spectroscopy remains largely theoretical and is only for homogeneous case. In actual cold atom experiments, there is usually a spatial confinement, mostly a harmonic trapping potential. The influence of such a confinement potential and the validation of quench spectroscopy methodology in such a situation, remains unknown.

In this work, we carry out the quench spectroscopy experimentally for one-dimensional ultracold bosons in the presence of optical lattices and harmonic trap confinement. From the measured momentum distribution evolution after quench, we apply Fourier transform and obtain the quench spectral function (QSF) $S(k, \omega)$ , which contains detailed information of the energy spectrum. For the MI case, we quench to the same final MI phase from different the initial lattice depths and find a modified signal of the Mott gap. More precisely, we detect a broadening of the bands' signal and a momentum cutoff at the half first Brillouin zone, which can be explained by the additional excitations caused by the harmonic confinement. Nevertheless, we can still distinguish the QSF from the one of SF phase where no gap is detected. All these observed properties are in good agreement with the DMRG simulations qualitatively.  Furthermore, with statistical analysis, we estimate the gap visibility for our measurements and find the parameters with best visibility. 
Our work reveals the physical mechanism behind the modified QSF of confined 1D ultracold boson systems in optical lattice, and explores the optimal conditions for applying quench spectroscopy.

\lettersection{Model and approach}
\label{sec.II}

The dynamics of the system can be described by the time-dependent Schr\"{o}dinger equation
\begin{eqnarray}
    i\hbar\frac{\partial}{\partial t}\psi\left(x,t\right)=\hat{H}\psi\left(x,t\right).
    \label{eqn_1}
\end{eqnarray}
In this work, we consider a one-dimensional boson gas (Lieb-Liniger gas) in the presence of external potentials, which Hamiltonian $\hat{H}$ writes
\begin{eqnarray}
    \hat{H}&=&\sum_i\left(-\frac{\hbar^2}{2m}\frac{\partial^2}{\partial x_i^2}+V\left(x_i, t\right)\right)+g\sum_{i<j}\delta\left(x_i-x_j\right)\nonumber\\
    \label{eqn_2}
\end{eqnarray}
with $m$ the mass of the particle, $x_i$ the position of the $i^{th}$ particle and $g$ the 1D coupling constant decide by the 3D scattering length $a_s$ and the transverse confine potentials. The external potential consists of two parts. It writes
\begin{eqnarray}
    V\left(x, t \right)=V_x(t) \sin^2{\left(kx\right)}+\frac{1}{2}m\omega^2x^2
    \label{eqn_3}
\end{eqnarray}
The first term is the periodic lattice potential along $x$ direction, with $k=\pi/a$ the wave vector of the lattice, $a$ the lattice period and $V_x(t)$ the lattice depth which is quenched from $V_x(t=0)=V_x^i$ to $V_x(t>0)=V_x^f$ in the actual experiment. The second term is the harmonic  potential $V_{HT}$ induced by lattice beams along the transverse $y-z$ directions, with a trapping frequency $\omega$.

\begin{figure*}
    \centering
    \includegraphics[width=1\linewidth]{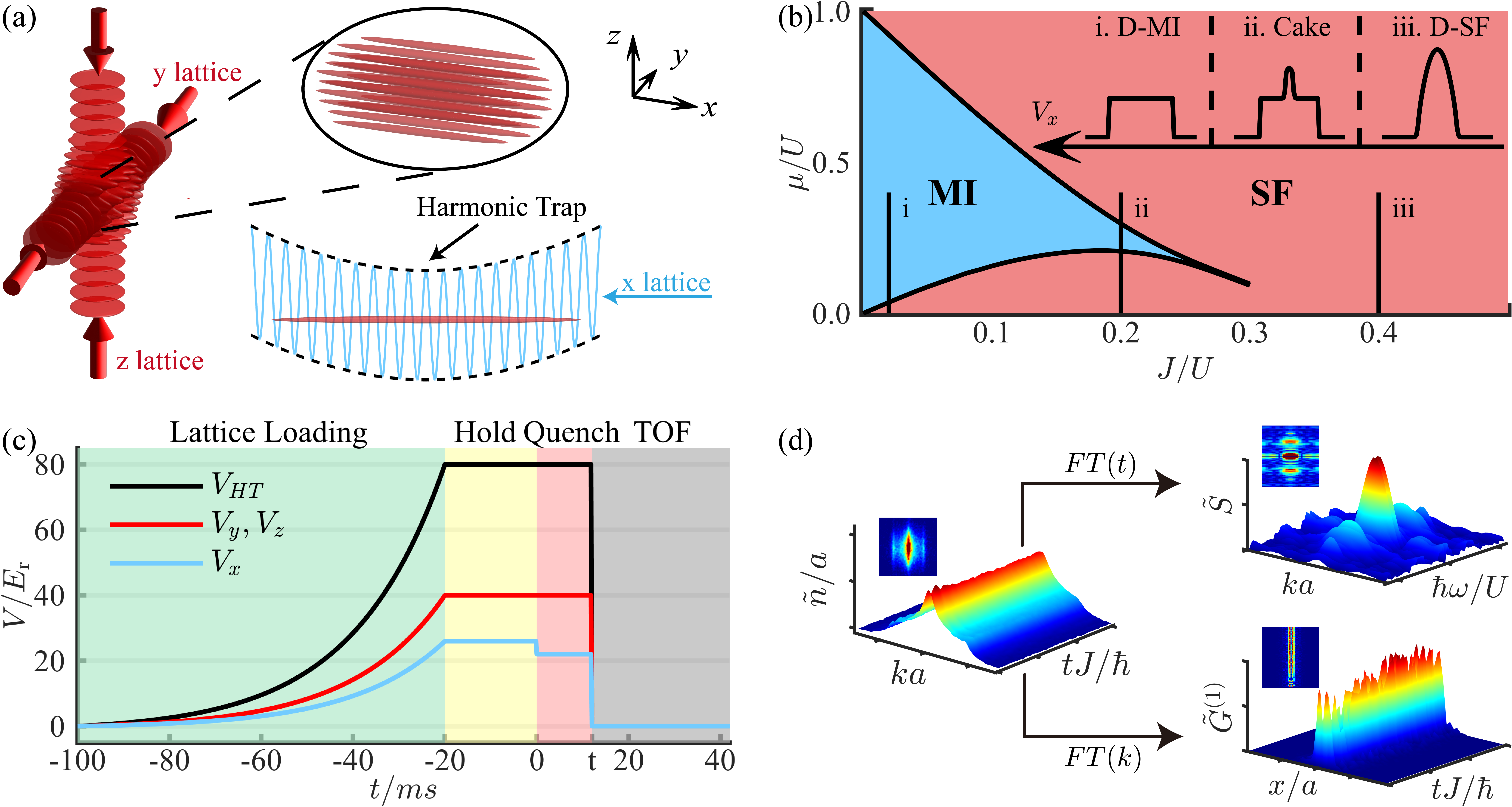}
    \caption{The quench spectroscopy experiment. (a) Sketch of the experimental setup. With two strong lattice beams($y$ and $z$ directions, red arrow), we form a bunch of one-dimensional atomic tubes (red arrays in set). Along the third direction, a relatively weak lattice ($x$ direction, blue) is added, on top of the presenting harmonic trap (black dashed lines) induced by the $y$-$z$ lattice beams.
    (b) Phase diagram for one-dimensional Bose-Hubbard model around the Mott lobe $na=1$. Superfluid (SF, red) and Mott insulator (MI, blue) phases are found at weak and strong interaction regimes, correspondingly. 
    (i)-(iii) represent the three regimes we focused, as we changing the lattice depth along $x$ direction $V_x$. The typical density distribution in real space are presented in the inset pictures.
    (c) The sequence of quench spectroscope. Below $t=0$ ms, we load the transverse lattices ($y$ and $z$ directions, red) and longitudinal lattices ($x$ direction, blue) and hold the system for $20$ ms. Correspondingly, a harmonic trap potential $V_{HT}$ is induced (black). At $t=0$ ms, we quench $V_x$. Then, all lasers are shut down at $t=20$ ms and we perform the TOF detection.  
    (d) Demonstration of data processing. From the TOF detection, we measure the momentum distribution evolution $n(k,t)$(left). By performing Fourier transform (FT) for parameter $t$ or $k$, we obtain the QSF $S(k,\omega)$(right top) or correlation spreading $G^{(1)}(x,t)$, correspondingly. The insets show the top view.
 }
    \label{fig:Fig.1}
\end{figure*}

Our experiment starts from preparing a nearly pure $^{87}$Rb BEC with atom number about $8.0 \times 10^4$, scattering length $a_s=100.4a_0$ and temperature below $30~{\rm nK}$. Notably, the finite temperature only influences the relative intensity of the excitation spectrum and does not shift the energy positions of the spectral peaks. (see more details in the Supplementary Material~\ref{SM sec.4}). Then we load the BEC into a three-dimensional optical lattice formed by three pairs of orthogonal retro-reflected laser beams. As illustrated in Fig.\ref{fig:Fig.1}(a), an array of one-dimensional tubes along $x$ direction is generated by two strong lasers along $y$ and $z$ directions, forming 2D optical lattices with lattice depth $V_y=V_z=40\Er$, with an uncertainty $5\%\sim10\%$. The weighted average atom number in each tube is about $\bar{N}=45$. We further add one-dimensional optical lattice $V_x$ (blue) along the tubes to carry out the quench spectroscopy. Due to the longitudinal distribution of the $y-z$ lattice beams, a Gaussian profile is superposed to the $x$ lattice, forming a harmonic trap in the center (black dashed line). 

Fig.\ref{fig:Fig.1}(b) gives the phase diagram of the corresponding Bose-Hubbard model without the trap $\hat{H}=-J\sum_i\left(\hat{a}_i^{\dagger}\hat{a}_{i+1}+\rm{H.c.}\right)+\frac{U}{2}\sum_i\hat{n}_i\left(\hat{n}_i-1\right)-\mu\sum_i\hat{n}_i$, with $J$ the tunneling between nearest-neighbor, $U$ the on-site interaction and $\mu$ the chemical potential~\cite{boeris-1dshallow-lattice-2016, villa2019, haller-pinning-2010}. The Mott insulator lobe with filling $na=1$ appears in the strongly-interacting regime, surrounded by superfluid phase. 
Due to the existence of the harmonic trap, the system can exhibit different phases at different sites. With the DMRG simulations (see details in the supplementary material), we find three typical density distributions, see inset of Fig.\ref{fig:Fig.1}(b). At low enough potential, the system is an inhomogeneous superfluid and we name it deep superfluid (D-SF) regime. Increasing the potential above $V_{c1}=9\Er$, we find Mott plateaus appear and form the typical wedding cake shape with a superfluid region $na>1$ in the middle. Further increasing the potential above $V_{c2}=21\Er$, the superfluid area with filling $na>1$ even disappears and the central part remains a large Mott plateau, which we name it deep Mott regime (D-MI). The three typical cases correspond to the three cuts (i-iii) in Fig.\ref{fig:Fig.1}(b). According to this phase diagram, we select the final state of the quench to make sure the system is quenched into the D-MI or D-SF regimes.

The time sequence of our experiment is shown in Fig.\ref{fig:Fig.1}(c). We load the $y-z$ lattices $V_y=V_z=40 \Er$ to form 1D tubes, which also causes the harmonic trap potential along the tubes $V_{HT}$. Correspondingly, the trap frequency of $V_{HT}$ is $\omega=2\pi\times57.9~\rm{Hz}$ with an uncertainty below $5\%$. We also load the $x$ lattice to the initial state of quench $V_x=V_x^i$ adiabatically in $80~{\rm ms}$, and then hold the system for $20~{\rm ms}$ to reach equilibrium. At the moment $t=0~{\rm ms}$, the lattice depth $V_x$ is suddenly changed to the final state $V_x^f$, and then the system evolves for a time $t_e$. Finally, we shut down all the lattices and, after $30~{\rm ms}$ ballistic expansion, we conduct the time-of-flight(TOF) detection, the imaging resolution of our system is $6.45~\rm{\mu m}$. In Fig.\ref{fig:Fig.1}(d), we show one typical example of the data we obtain, for the case we quench from $V_x^i=18 \Er$ to $V_x^f=22 \Er$. For the evolution time $t_e$ varying from $0$ to $12~{\rm ms}$, we can get the momentum distribution $n(k_x,k_y,t)$ from TOF imaging along $z$ direction. By integrating along $y$ direction,  we obtain the time-evolution of the 1D momentum distribution $n(k_x,t)$. As suggested by Refs\cite{villa2019, villa2021a, villa2021b}, we can apply Fourier transform to the $k$ or $t$ variables, such that we can obtain the spreading of one-body correlation function $G^{(1)}(x,t)$ or the QSF $S(k, \omega)$, correspondingly. Noted that due to the breaking of translational invariance in the presence of $V_{HT}$, $G^{(1)}(x,t)$ here represents the integrated correlation function which can be written as $G^{(1)}(x,t)=\int dx' g^{(1)}(x'+x,x')$ with $g^{(1)}(x'+x,x')=\left\langle\hat{\psi}^\dagger\left(x'+x,t\right)\hat{\psi}\left(x',t\right)\right\rangle$ the one-body correlation function for the homogeneous case. For details about data processing steps, see Supplemetary Material \ref{SM sec.3}.

Numerically, we simulate this model using the Density Matrix Renormalization Group (DMRG) method\cite{white1992,schollwock2011} and Time Evolving Block Decimation (TEBD) algorithm based on TeNPy project\cite{tenpy2024}. For the initial state, we apply the particle number operator at different positions to extract particle number distribution $n(R)$. For each step of time evolution after quench, we apply the creation and annihilation operators at different positions and extract the one-body correlation function $G^{(1)}(x,t)$. Then, we can compute the momentum distribution $n(k,t)$ and the QSF $S(k,\omega)$ by applying Fourier transform. 
In the actual simulation, we use $L=151a$ with open boundary condition, and various particle numbers averaged over different tubes in accordance with experiment(see below and Supplementary Material~\ref{SM sec.5}).

\lettersection{Quench spectroscopy for different phases}
\label{sec.III}

\begin{figure}
    \centering
    \includegraphics[width=1\linewidth]{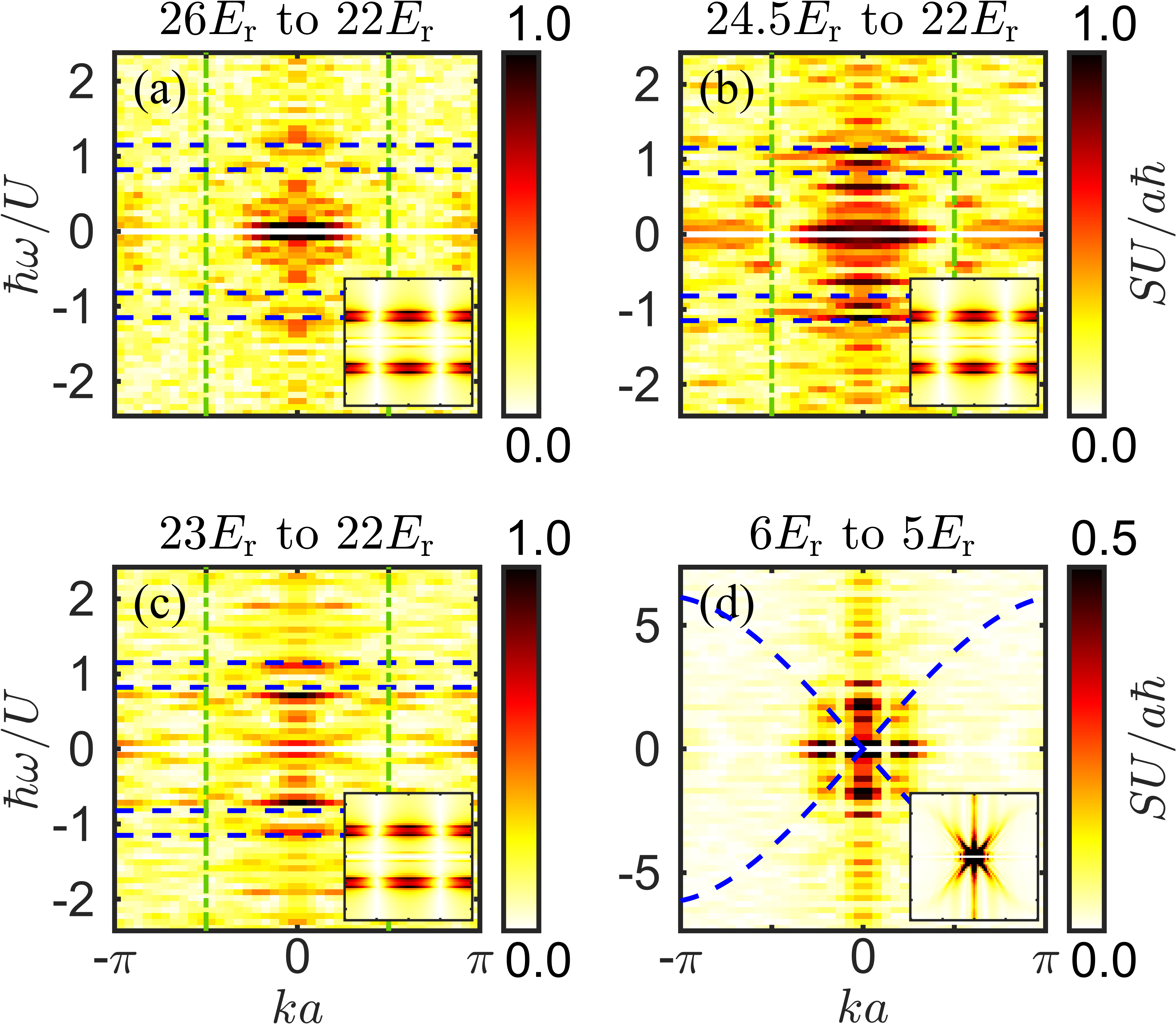}
    \caption{QSF measured from the experiment when quenching from different initial trap depth $V_x^i$ to a lower final depth $V_x^f$. (a). $V_x^i=26\Er$ to $V_x^f=22\Er$, (b) $V_x^i=24.5\Er$ to $V_x^f=22\Er$, (c) $V_x^i=23\Er$ to $V_x^f=22\Er$, (d) $V_x^i=6\Er$ to $V_x^f=5\Er$. (a)-(c) are in the deep MI regime and (d) is in the deep SF regime. The values in each figures result from the average over $6$ sets of experimental data, and the average relative error is around $20\%$ (details of error analysis, see Supplementary Material \ref{SM sec.6}). 
    Inset figures represent the corresponding DMRG simulations. 
    In (a)-(c), the blue and green dashed lines are guidance for the eyes to view the broadened peaks and momentum cutoff suggested by the simulation. 
    The blue dashed curves in (d) are Bogoliubov spectral branch suggested by the simulation result. The X-axis represents $k$ rescaled by $1/a$, Y-axis is $\omega$ rescaled by $U/\hbar$. The Colorbars represent $S(k,\omega)$ rescaled by $a\hbar/U$.}
    \label{fig:Fig.2}
\end{figure}

Firstly, we focus on the detection of the gap signal for the MI phase. With the presence of harmonic trap, it is not guaranteed that the information of the Mott gap still remains clearly in the QSF. Therefore, we firstly fix the final lattice depth $V_x^f= 22\Er$ in the D-MI domain, and probe the QSF under different quench amplitudes. More precisely, we set the initial lattice depths $V_x^i$ to $26\Er$, $24.5\Er$ and $23\Er$ respectively. For each condition, the evolution time after quench $t_e$ is gradually increased to $12~{\rm ms}$ at a time interval of $0.2~{\rm ms}$. Under this condition, the maximum frequency we can detect is $\omega_{max}\approx2.42U/\hbar$ with a resolution $\Delta\omega\approx0.08U/\hbar$, which is sufficient for our observations. Right after the quench stage, we obtain the momentum distribution evolution $n(k_x,t)$ by a $30~{\rm ms}$ TOF detection. 
As illustrated in Fig.\ref{fig:Fig.1}(d), we calculated the corresponding QSF and the results are shown in Fig.\ref{fig:Fig.2}.

In Fig.\ref{fig:Fig.2}(a)-(c), we show the amplitude of measured QSF for different $V_x^i$ to $V_x^f = 22\Er$. The insets give the DMRG results for the corresponding experimental parameters. The DMRG simulation suggests that an energy gap $\Delta \sim U$ can be observed. However, instead of a sharp line as in the homogeneous case, this signal has a broadening of the line width $w_{DMRG}(\hbar \omega=U) = 0.164 U$.  
This can be understood by the non-degenerate nearest-neighbor tunneling excitations inside the Mott plateau with the presence of an inhomogeneous trap. 
In practice, it provides an energy shift of $\Delta V(i) =\pm \big(V_{HT}(x_{i+1})-V_{HT}(x_i)\big) = \pm m\omega^2a^2(2i+1)/2$ for the excitation between the $i-\mathrm{th}$ and $(i+1)-\mathrm{th}$ site, leading to a line width $w(\hbar \omega=U) = m\omega^2a^2(2i_{max}+1)$. According to the DMRG simulation of the density profile, we have $i_{max} =21$ which is about half width of the Mott plateau. It gives $w(\hbar \omega=U)= 0.170U$ which fits with our numerical simulation within $4 \%$. In our experimental data, we also detect such a broadened signal around $\hbar \omega = U$, with width $w_{exp}(\hbar \omega = U)=0.212\pm0.023$. This fits with our theoretical prediction within $25 \%$. Notably, the signals at $\hbar \omega = 0$ is also broadened into an ellipse shape, which is probably due to the tube distributions in actual experiment, and this is confirmed in our further DMRG calculation, see Supplementary Material \ref{SM sec.5}. Besides, the signal is strongest for an intermediate quench ($V_z = 24.5E_r \rightarrow 22E_r$), reflecting a balance between too small quenches, which produce only weak deviations from equilibrium, and too large ones, which overpopulate high-energy excitations and suppress the lowest branch. This observation provides useful insights for optimizing quench spectroscopy.

Moreover, both the experimental and the DMRG data shows that the quench spectral function exhibits a clear suppression around $ka=\pi/2$, whereas a homogeneous system remains continuous in $k$. 
This effect can be attributed to the interference of two symmetric nearest-neighbor excitation, which is proportional to
$\langle n | \hat a_k^\dagger \hat a_k | 0 \rangle \,\delta(E_n-\hbar\omega)$, with $|n\rangle$ exicited state.
In the Mott regime the relevant excitations are nearest-neighbor doublon-holon pairs such as
$\quad\hat a_{-(R+1)}^{\dagger}\hat a_{-R}|g\rangle\quad$and
$\quad\hat a_{R+1}\hat a_{R}^{\dagger}|g\rangle\quad$,
which possess the same excitation energy and therefore interfere. Substituting these states yields a $\cos(ka)$ dependence of the matrix element, which vanishes at $k=\pi/2$ and accounts for the observed suppression. 
Furthermore, for $k>\pi/2$ high quasi-momentum modes are particularly sensitive to phase coherence, so experimental decoherence further reduces the observable spectral weight.

Even though the harmonic trap tends to blur the information of the spectrum as discussed above, we can actually still distinguish the MI phase from the SF phase. To further prove this, we do the same experiment with the quench set from $V_x^i = 6\Er$ to $V_x^f = 5\Er$, see Fig.\ref{fig:Fig.2}(d) with the inset the corresponding DMRG calculation. Under this condition, the system locates in the deep SF regime.
In this picture, at $ka=0$, we can clearly see a continuous distribution in both the experimental data and numerical simulations, which is the signature of gapless properties for the SF regime. 
By plotting a cut at $ka=0$ and $ka=\pi/5$ for both of the four cases, we can clearly distinguish the gap and gapless signature between MI and SF phase, even with the presence of the harmonic trap. Notably, in the simulation data for the case in Fig.\ref{fig:Fig.2}(d) , we can also see a spectral branch extending from $S(0,0)$ (see the and blue dashed line which is from the light yellow structure in inset), which agrees with the prediction based on Bogoliubov theory\cite{pitaevskii2004}. The strength of this spectral is only $2 \%$ of the maximum at $k$ far from zero, which can not be fully observed in actual experimental data due to the weakness of this signal.

\lettersection{Inverse quench and statistical analysis}

\begin{figure}
    \centering
    \includegraphics[width=1\linewidth]{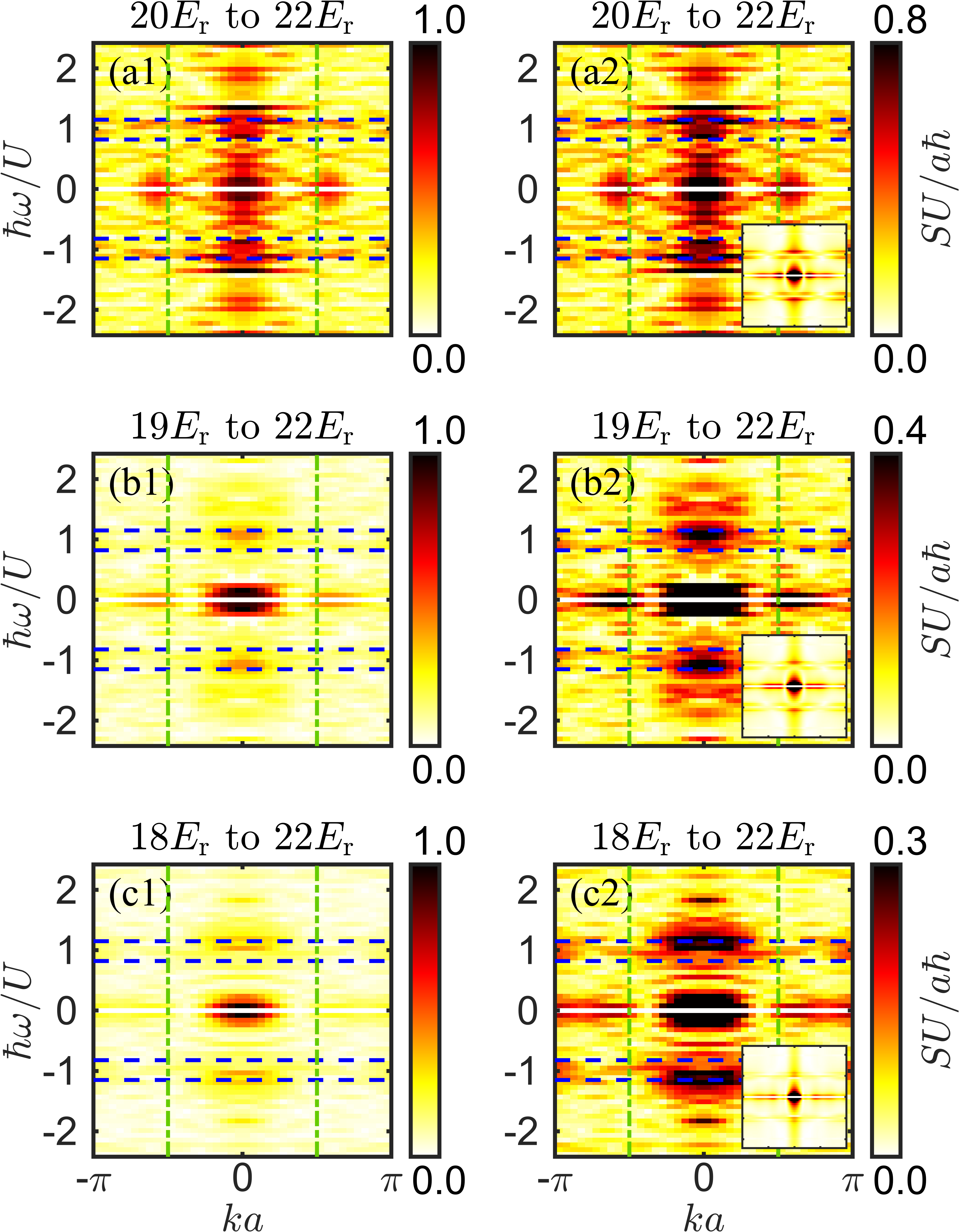}
    \caption{Amplitude of QSF when quenching from different $V_x^i$ to a higher $V_x^f=22\Er$ in the deep MI phase. (a)-(c) are cases quenching from different $V_x^i=20E_r,19E_r,18E_r$ to a fixed $V_x^f=22E_r$ respectively.(a1)-(c1) shows the original QSF with the same range of color bar, while (a2)-(c2) adjust the color bar to reveal the detailed structure of the QSF. The values in each figures result from the average over $6$ sets of experimental data, and the average relative error is around $20\%$. Inset figures represent the corresponding DMRG simulations. The blue and green dashed lines are guidance for the eyes to view the broadened peaks and momentum cutoff suggested by the simulation. The axis and Colorbars represent the same meaning as in Fig.\ref{fig:Fig.2}.
    }
    \label{fig:Fig.3}
\end{figure}

To seek for better signal of the spectrum of the D-MI phase, we further test the inverse quench process, namely $V_x^i < V_x^f$. The final state is fixed at $ V_x^f=22\Er$, while the initial trap depths are set to $V_x^i = 20\Er$, $19\Er$ and $18\Er$, see Fig.\ref{fig:Fig.3}. We measure the QSF while keeping the other experimental parameters same as Fig.\ref{fig:Fig.2}(a-c). The first row in Fig.\ref{fig:Fig.3} is the results normalized to $S_{max} U/a\hbar =1$. At the zero momentum cut $ka=0$, a clear gapped structure is observed for the case $V_x^i = 19\Er$ and $18\Er$. However, the high momentum signal is weak. By setting the proper cutoff to the colorbars of each plots, we improve the visibility of the information inside, see Fig.\ref{fig:Fig.3}. (a2)-(c2). The inset figures are the corresponding DMRG results.

\begin{figure}
    \centering
    \includegraphics[width=1\linewidth]{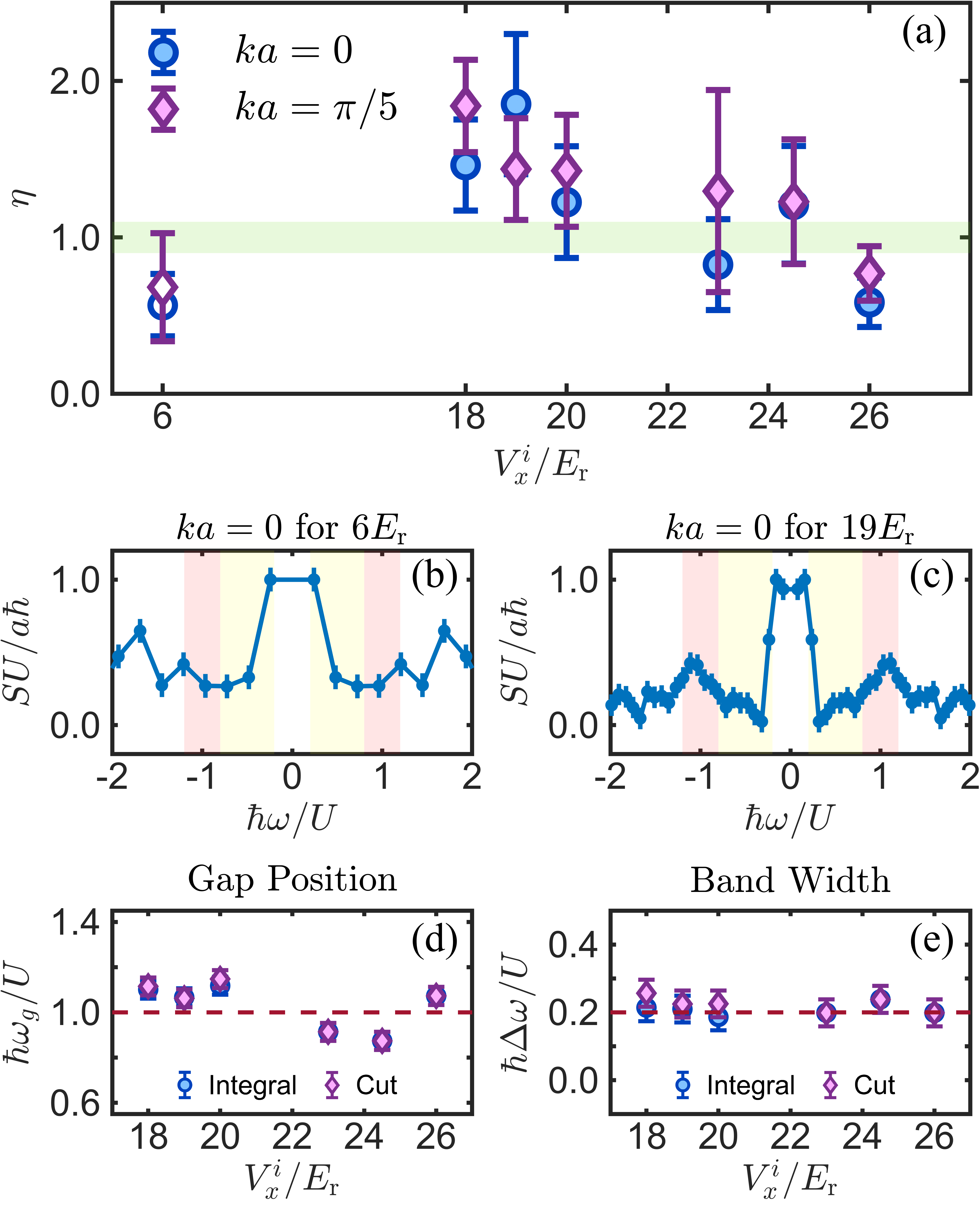}
    \caption{Visibility of the energy gap. (a) $\eta$ for different experiments. The blue and purple data points are calculated at cutting $ka=0$ and $ka=\pi/5$, respectively. The light green area are the line $\eta=1\pm 0.1$ to guide the eyes. (b) cutting at $ka=0$ for SF quench. (c) cutting at $ka=0$ for MI quench from $19\Er$ to $22\Er$. The red and yellow region in (b) and (c) represent the spectral area and gap area used to calculate $\eta$. (d) Position of the energy gap. The red dash line shows the theory position $\hbar\omega/U=1$; (e) Bandwidth of the energy gap. The blue and purple data points in (d) and (e) is calculated according to integral and cut of $S(k,\omega)$, respectively.
    }
    \label{fig:Fig.4}
\end{figure}

From these data, we can conclude that with the decreasing of initial trap depth, the relative strength of the spectral branch also decrease, but the structure of spectral is even clearer, especially the high momentum part. Moreover, comparing with the quench with decreasing lattice depths, we find this inverse quench tends to give a better spectrum for our system. 
We attribute this to the fact that with a smaller initial potential, the corresponding equilibrium phase of $V_x^i$ enters the cake regime where more superfluid component enters into the density distributions with the presence of harmonic trap and the width of the Mott plateau also decreases, see Supplementary Material \ref{SM sec.2}. 
This causes additional gapless excitations around $\hbar\omega=0$ due to superfluid-Mott tunneling,  and clearer Mott gap excitation signals since the width of Mott plateau decreases. Notably, similarly as the cases in Fig.\ref{fig:Fig.2} (a-c), the band at $\hbar\omega=U$ has a width $w(\hbar \omega=U)=0.203\pm0.015$, which fits the theoretical estimations within $20 \%$. The main signal of these bands also ends around $k=\pi/2$ as predicted before.

We further evaluate the visibility of the energy gap in all the obtained QSF above. 
According to the previous discussions, the Mott insulator we detect has an energy gap $U$ with a blurred width $w=0.2U$. Thus we define the gap visibility $\eta$ as 
\begin{eqnarray}
    \eta=\frac{\bar{I}_s}{\bar{I}_g}=\frac{\displaystyle \frac{1}{2w}\int_{U-w}^{U+w}S\left(k,\omega\right)d\omega}{\displaystyle \frac{1}{U-2w}\int_{w}^{U-w}S\left(k,\omega\right)d\omega}
    \label{eq_vis}
\end{eqnarray}
with $\bar{I}_s$ the average intensity within the area $[\pm U-w, \pm U+w]$ and $\bar{I}_g$ the average intensity of $[w,U-w]$. Fig.\ref{fig:Fig.4}(a) gives $\eta$ at $ka=0$ and $ka=\pi/5$ for all different experimental parameters in Fig.\ref{fig:Fig.2} and Fig.\ref{fig:Fig.3}. The data at $V_x^i=6\Er$ represent SF quench from $6\Er$ to $5\Er$, and the others are MI quench from different $V_x^i$ to $22\Er$. 
The light green area are the line $\eta=1\pm 0.1$ to guide the eyes.
From the figure, we can see that for the SF case, $\eta(V_x^i=6\Er)$ is clearly less than $1$. While for the MI cases, $\eta$ is larger than $1$ for most of the cases. Especially, for the cases $V_x^i=18 \Er$ and $19 \Er$, $\eta$ is the largest and clearly more than the threshold $1+10\%$ we set. This fits with our previous statement that the case of the inverse quench with slightly larger quench amplitude shows the most obvious gap signals. Their QSF can be easily distinguished from the SF one. 

Furthermore, we quantitatively evaluate the energy gap by estimate the position and width of the excited band. For the measured $S(k,\omega)$, we either take a cut of  at $ka=\pi/5$  or perform an integral between $ka=0$ and $ka=\pi/2$, and
evaluate the excited band position $\omega_g$ and width $\Delta \omega$, see Fig.\ref{fig:Fig.4}(d) and (e) correspondingly~(see more details in Supplementary Material \ref{SM sec.6}).
From the integral method (blue circles), the average band position (equivalently gap width) is $\hbar\omega_g/U=1.024\pm0.104$ and the average bandwidth is $\hbar\Delta\omega/U=0.208\pm0.018$.From the cut method (purple diamonds), the average excited band position is $\hbar\omega_g/U=1.031\pm0.111$ and the average bandwidth is $\hbar\Delta\omega/U=0.224\pm0.022$.
They both fit with our conclusion that a gap is observed at $\hbar\omega_g/U \simeq 1\pm0.1$.

\lettersection{Conclusion}
In conclusion, we implement a quench spectroscopy for one dimensional Bose-Hubbard model in the presence of harmonic trap. We observe a spectral branch at the position predicted theoretically. In contrast to homogeneous system, the spectral is a broadened area rather than a sharp line due to the existence of the harmonic trap. And the spectral has a cutoff at $ka=\pi/2$ at the same time. We attribute these phenomenon to the non-degenerate nearest-neighbor excitations caused by the harmonic confinement. Moreover, we study the performance of the quench spectroscopy and compare spectral visibility of various quenches. We can see an obvious distinction between the two phases, and we find that we can obtain a clearer spectrum when we perform an inverse quench with slightly larger quench amplitude. 

Our research verifies the validity of quench spectroscopy for detecting the excitation spectrum of many-body quantum phases of 1D ultracold bosons in optical lattice with the presence of a potential confinement. In addition to simulate confined 1D Bose-Hubbard Model , it could also potentially be applied to the research of other models, such as Bose glass phase in disordered~\cite{Smith_NP_2016}\cite{villa2021a}\cite{villa2021b} or quasiperiodic systems~\cite{gautier-2Dquasicrystal-2021,yao-boseglass-2020}, dipolar XY model\cite{quench_XY_Science_Antoine_Browaeys}, quantum spin chain\cite{quench_quantum_spin_prb, Schneider_PRR_2021},and dissipative non-Hermitian quantum lattice models\cite{quench_dissipative_non_Hermitian}.

\lettersection{Acknowledgments}
We thank Laurent Sanchez-Palencia and Botao Wang for helpful discussions. This research was supported by the National Key Research and Development Program of China (Grants No. SQ2021YFA1400224, No. 2021YFA0718300, No. 2021YFA1400900), the National Natural Science Foundation of China (Grant No. 11920101004) and The Fundamental Research Funds for the Central Universities, Peking University.

\lettersection{Data availability}
The data that support the findings of this article are openly available at \cite{dataset}.

\bibliographystyle{revtex}
\bibliography{biblio-HY}


 \renewcommand{\theequation}{S\arabic{equation}}
 \setcounter{equation}{0}
 \renewcommand{\thefigure}{S\arabic{figure}}
 \setcounter{figure}{0}
 \renewcommand{\thesection}{S\arabic{section}}
 \setcounter{section}{0}
 \onecolumngrid  
     
 
 \newpage

 {\center \bf \large Supplemental Material for \\}
 {\center \bf \large Quench spectroscopy for Lieb-Liniger bosons in the presence of harmonic trap  \\ 
 }

\vspace*{1.cm}

In this supplemental material, we provide details about the DMRG method, the density distribution in harmonic trap, processing methods and analysis of experimental data and the effect of finite-temperature and atom number distributions over different tubes.

\section{The DMRG method}
\label{SM sec.1}
The numerical results in this paper are obtained by density-matrix renormalization group approach (DMRG), based on the TeNPy project~\cite{tenpy2024}.
The approach resorts on singular value decomposition (SVD) to decompose quantum states into matrix product states (MPS), which squeezes the Hilbert space while keeping the most relevant information.  To simulate the preparation of the initial state, we use two-site DMRG algorithm to calculate the ground state, and then TEBD algorithm is applied to simulate the time evolution after the quench.

In our case of finite-size 1D Bose-Hubbard Model with harmonic trap, the reduced Hilbert space dimension is determined by truncation of local Hilbert space $n_{max}$ and bond dimension $\chi_{max}$ between the sites. To validate the accuracy of our simulation, we follow the methodology of Ref.~\cite{villa2019}, choosing $n_{max}=5$, $\chi_{max}=200$ for Mott insulator (MI) regime  and $n_{max}=10$, $\chi_{max}=500$ for superfluid (SF) regime. We also perform the relevant convergence tests to prove our choice is sufficient. Below we give the details of the truncation.

\vspace{1\baselineskip}

\textit{Truncation of local Hilbert space}

\begin{figure}[t!]
    \centering
    \includegraphics[width=0.88\linewidth]{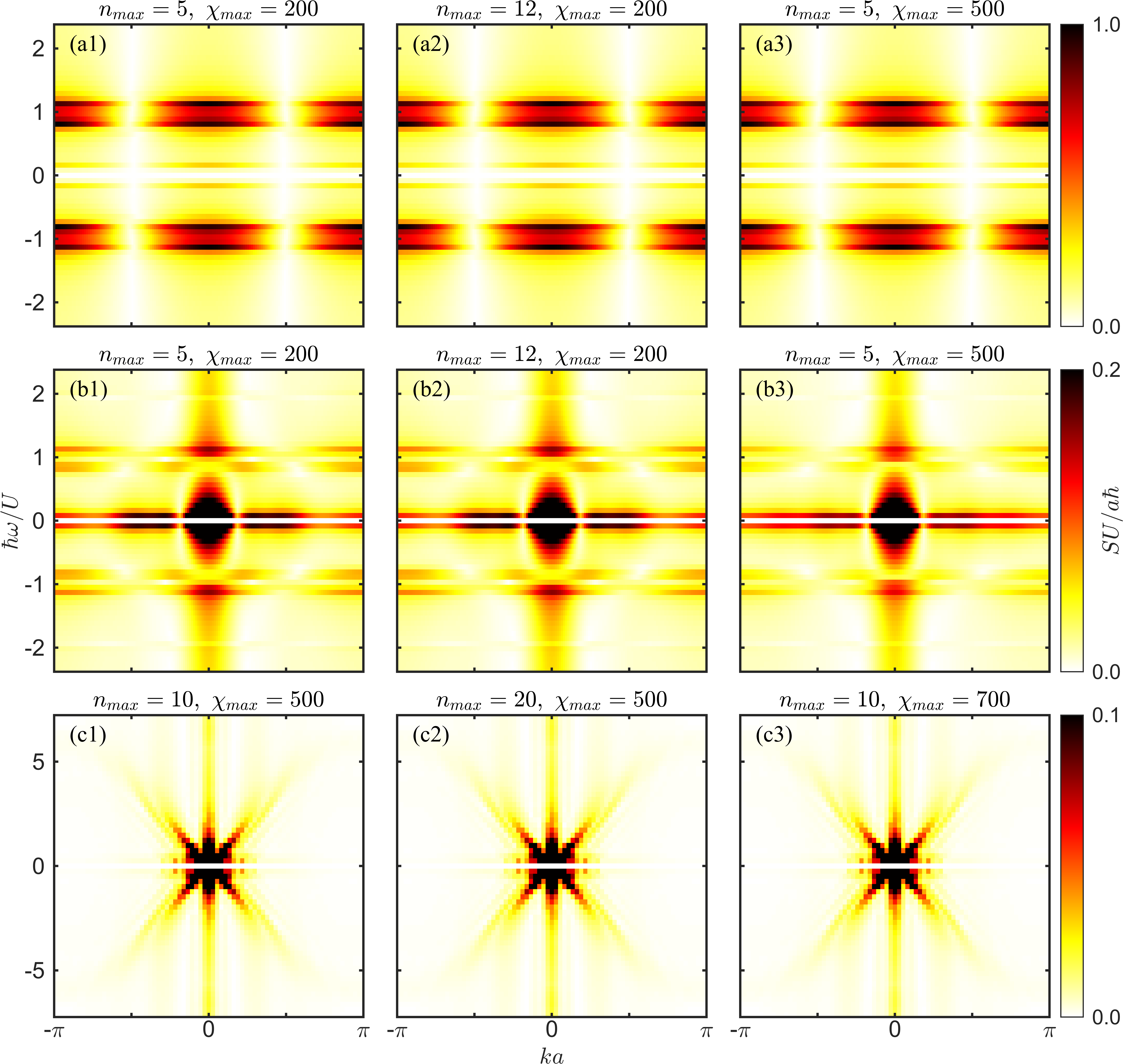}
    \caption{Amplitude of QSF for three simulations of different typical regimes under original and higher $n_{max}$ and $\chi_{max}$, where (a) is from $V_x=24.5Er$ to $V_x=22Er$, (b) is from $V_x=19Er$ to $V_x=22Er$ and (c) is from $V_x=6Er$ to $V_x=5Er$. And parameters $U$ and $J$ are calculated using the s-wave scattering length $a_s=100.4a_0$, where $a_0$ is Bohr radius.}
    \label{fig:Fig.S1}
\end{figure}

We follow the criteria $1-\sum_{n=0}^{n_{max}}P(n) \le 10^{-2}$, where $P(n)$ is the probability that n bosons occupy a given lattice site. We can set $P(n) = \bar{n}^ne^{-\bar{n}}/n!$, which is the Poisson distribution. This is approximately valid for SF regime while MI regime has a faster decay and less fluctuations. Thus, it is a good upper estimation. It can be examined that $n_{max}=5$, $\chi_{max}=200$ for MI regime and $n_{max}=10$, $\chi_{max}=500$ for SF regime fulfill this criteria. Additionally, we choose the typical QSF calculation in the main text and change $n_{max}$ to perform convergence tests (Fig.~\ref{fig:Fig.S1}). Indeed, as shown in the first and second columns of Fig.~\ref{fig:Fig.S1}, we see no significant difference of quench spectral function comparing with higher $n_{max}$.

\vspace{1\baselineskip}

\textit{Truncation of bond dimension}

Bond dimension is the matrix dimension of the MPS coefficients. Explicitly, the many-body wave function writes
$$\lvert\Psi\rangle = \sum_{n_1, n_2, \dots, n_M} A^{n_1}[1] A^{n_2}[2] \dots A^{n_M}[M] \lvert n_1, n_2, \dots, n_M \rangle$$, where the $A^{n_j}[j]$ is a $\chi_{j-1}\times\chi_{j}$ matrix, and the $\chi_{j}$ is the bond dimension between the $j$-th and $j-1$-th site. The truncation of bond dimension $\chi_{max}$ is the possible maximum of bond dimension, which affects the ability of MPS to keep the entanglement information. In the simulation, $\chi_{max}$ is chosen sufficiently large so that the truncation does not affect the results significantly. And as shown in the second and third columns of Fig.~\ref{fig:Fig.S1}, we see no significant difference of QSF comparing with higher $\chi_{max}$.

\section{Density distribution in the presence of harmonic trap}
\label{SM sec.2}

In the presence of a harmonic trap, the density distribution of the system at equilibrium can exhibit different phases, be it MI or SF, on different lattice sites.
The effective chemical potential at site $x$ can be described as $\mu(x)=\mu - V(x)$, with $\mu$ the real chemical potential and $V(x)$ the harmonic trap. Consequently, for a fixed periodic lattice potential $V_x$, the sites across the system are arranged on a vertical line in the phase diagram. 
To verify the different regimes in Fig.~1(b) of the main paper and identify the appropriate parameters in the following experiments, we compute the density distribution for the harmonically trapped systems at varying $V_x$.

In the deep lattice limit ($V_x  \gg \Er$), the tunnelling amplitude $J$ can be quantitatively obtained by solving exactly the Mathieu equation, while the on-site interaction energy $U$ is typically estimated by approximating the lowest band Wannier functions by Gaussian functions\cite{RevModPhys.80.885}. They write 
$$\frac{J}{\Er} \simeq \frac{4}{\sqrt{\pi}}(\frac{V_x}{\Er})^{3/4}e^{-2\sqrt{V_x/\Er}},\quad\frac{U}{\Er} \simeq4\sqrt{2\pi}\frac{\as}{\lambda}(\frac{V_x}{\Er})^{3/4},$$
where $\Er=h^2/2m\lambda^2$ is the recoil energy with $\lambda$ the wavelength of the laser and $m$ the mass of the trapped atoms. Therefore, increasing  $V_x$ results in a shift of the vertical line in the phase diagram to the left, which leads to different typical density distributions, as we have mentioned in Fig.\ref{fig:Fig.1}(b).

\begin{figure}
    \centering
    \includegraphics[width=0.95\linewidth]{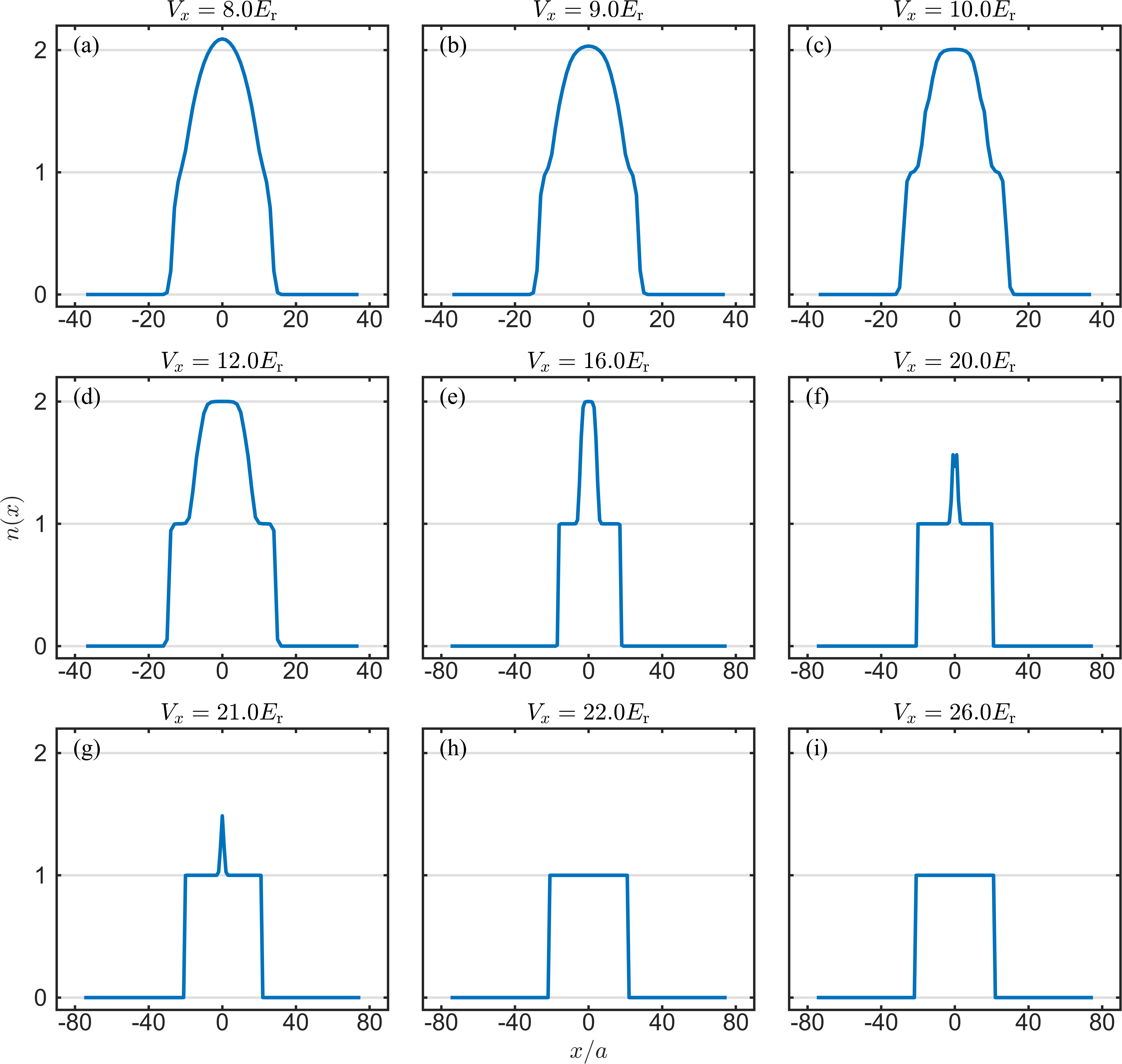}
    \caption{Density distributions for different $V_x$ calculated by DMRG under the conditions $L=151$, $N=45$, and $a_s=100.4a_0$}
    \label{fig:Fig.S2}
\end{figure}

At low enough $V_x$, the system is purely a superfluid with a space-dependent $\mu(x)$, see Fig.\ref{fig:Fig.S2} (a). When $V_x$ increases, the vertical line crosses the MI lobe with filling $na=1$, causing the system to shift into the cake region. The estimation of such a transition point $V_{c1}$ is possible through the execution of DMRG calculations, and a comparison of the various density distributions indicates that it should be around $V_{c1}=9\Er$ (see Fig.\ref{fig:Fig.S2} (a-c)).

Further increasing $V_x$, the top point of the vertical line moves below the upper boundary of the MI lobe, which annihilates the SF area with filling $na>1$ and remains only the Mott plateau in the center of the system. Comparing the density distributions in Fig.\ref{fig:Fig.S2}(f-h), we can estimate this transition point $V_{c2}\simeq21\Er$.

\section{Data processing steps and noise filtering}
\label{SM sec.3}

\vspace{1\baselineskip}

\textit{Data processing steps to extract $S(k,\omega)$}

First of all, for a certain evolution time $t_e$ (in our experiment, $t_e$ is from $0.2~\rm{ms}$ to $12~\rm{ms}$ at an interval ${\Delta}t=0.2~\rm{ms}$), we obtain the momentum distribution $n(k_x,k_y,t_e)$ by means of TOF detection. 
    
Secondly, we integrate the filtered $n(k_x,k_y,t_e)$ along $y$ direction, \ie we only keep the momentum distribution of $x$ direction, which is the direction of 1D Bose gas. Then we can get a row of data $n(k_x,t_e)=\sum_{k_y}n(k_x,k_y,t_e)$.
    
After that, by changing $t_e$ and repeat the above steps, we can obtain $60$ rows of data and build the data matrix $n(k,t)$. Each row of $n(k,t)$ represents the momentum distribution at the corresponding $t_e$.

Finally, for each \textbf{column} of $n(k,t)$, we perform a Discrete Fourier Transform(DFT) to extract $S(k,\omega)$, which can be writen as:

\begin{eqnarray}
    S(k,\omega)=\sum_tn(k,t)e^{-i{\omega}t}{\Delta}t
    \label{eqn_1}
\end{eqnarray}

\vspace{1\baselineskip}

\textit{The noise filtering of $n(k_x,k_y,t_e)$}

\begin{figure}
    \centering
    \includegraphics[width=0.5\linewidth]{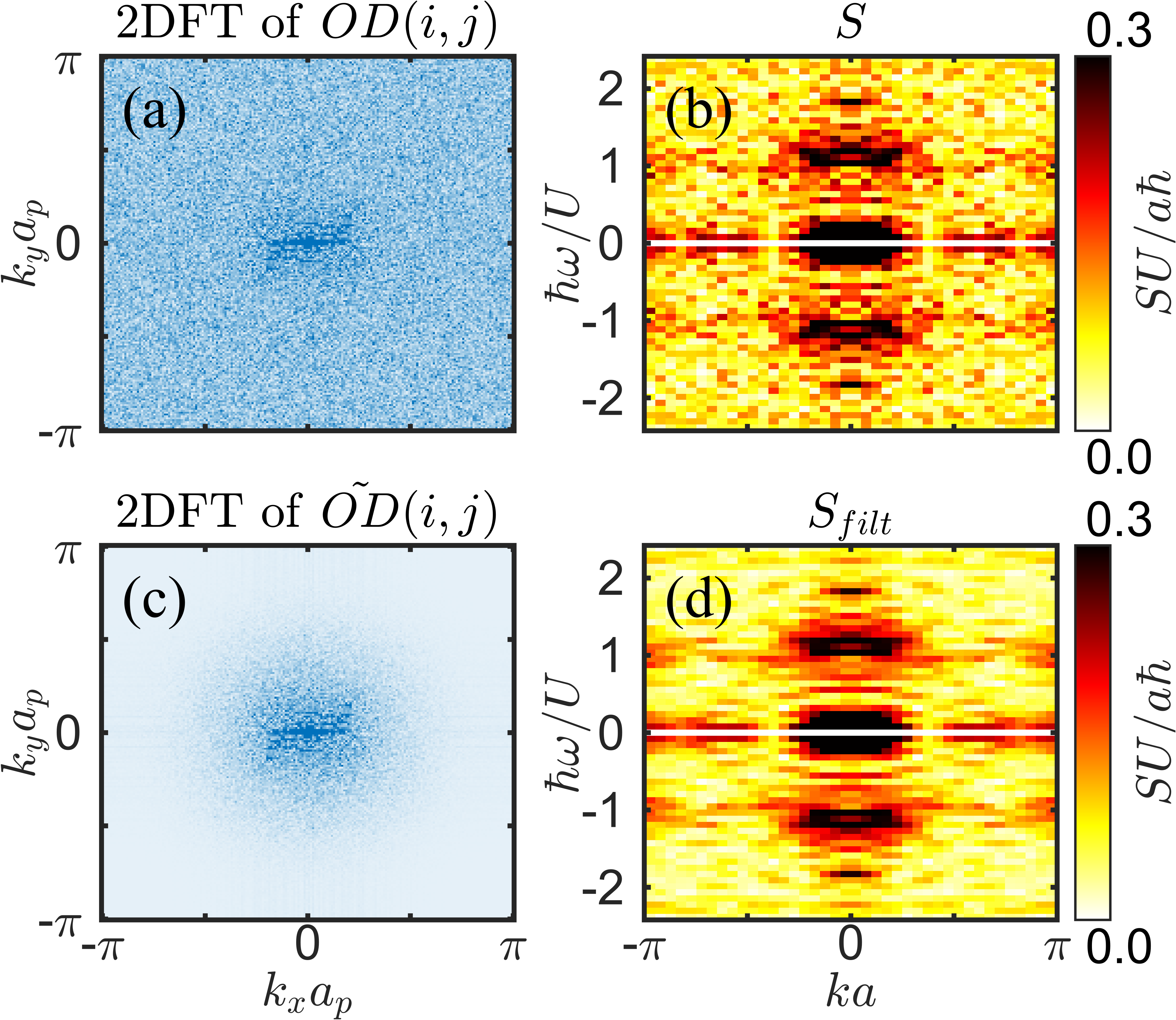}
    \caption{Comparison of $OD(i,j)$ and $S(k,\omega)$ with or without noise filtering. (a) 2DFT of origin $OD(i,j)$, (b) $S(k,\omega)$ before noise filtering, (c) 2DFT of filtered $\tilde{OD}(i,j)$, (d) $S(k,\omega)$ after noise filtering. The date used here is from quench $18 \Er$ to $22\Er$}
    \label{fig:Fig.S3}
\end{figure}

As mentioned above, to obtain the momentum distribution $n(k_x,k_y,t_e)$, we conduct TOF detection using a CCD camera. The pictures we get, which we denote as $OD(i,j)$, inevitably contain noises originating from shot noise, dark current and etc. In Fig.\ref{fig:Fig.S3} (a), we provide an example of the two-dimensional Fourier transform (2DFT) for $OD(i,j)$. The figure is taken for the inverse quench from $18\Er$ to $22\Er$ at $t_e=0.2~\rm{ms}$. We can see clearly that a noise background is mixed with the signal . On the other hand, the spectrum signal we obtain in the 2DFT is weak, see Fig.\ref{fig:Fig.S3}(b). That makes it difficult to read the information we need easily. Therefore, we implement image filtering to improve the signal-noise ratio, as done in Refs.~\cite{Img_Filt_Rev, GaussianFilter}. 

When we obtain the $OD(i,j)$ by means of TOF detection, we perform a Gaussian filtering to the average 6 sets of the experimental data. The standard deviation ($\sigma$) of the gaussian convolution kernel $G_\sigma$ is determined according to different regimes where the system locates in. For the MI and SF phases, $\sigma$ is set to $1.5$ and $1.0$, respectively. The corresponding sizes of the kernel are $7$ and $5$. Then we can get the filtered figures by $\tilde{OD}(i,j)=OD(i,j)*G_\sigma$, where $*$ represents convolution operation. Fig.\ref{fig:Fig.S3}(c) shows the filtered data of the one in Fig.\ref{fig:Fig.S3}(a). Clearly, the high spatial frequency noise is significantly suppressed. After this process, we rescale the $\tilde{OD}(i,j)$ by $Na$ to obtain the normalized $n(k_x, k_y,t_e)$, with $N$ the atom number and $a$ the lattice constant. Fig.\ref{fig:Fig.S3}(d) gives the QSF $S(k,\omega)$ obtained after the filtering process. Comparing with Fig.\ref{fig:Fig.S3}(b), a cleaner structure of the QSF can also be observed. 

\section{Finite-temperature effects on the quench spectroscopy signal}
\label{SM sec.4}
Here we provide a more formal analysis of how a finite-temperature initial state affects the excitation spectrum. 
For quench spectroscopy, the excitation spectrum is not destroyed by finite temperature. This is because the spectral response is induced by the overlap between the initial state and the excited eigenstates of the post-quench Hamiltonian. Starting from a thermal state, this overlap remains non-vanishing and may even be enhanced due to the population of low-lying excited states. The effect of temperature is therefore expected to modify mainly the relative intensities of spectral features, but not their energy positions.

This can be illustrated more formally by considering the initial thermal density matrix,  
\[
\hat\rho_i=\frac{1}{Z_i}\sum_m e^{-\beta_i E_m}\, |m\rangle_i \langle m|_i ,
\]  
where $|m\rangle_i$ are eigenstates of the pre-quench Hamiltonian with eigenenergies $E_m$. Expanding these states in terms of the eigenbasis $\{|n\rangle_f\}$ of the post-quench Hamiltonian,  
\[
|m\rangle_i = \sum_n U_{mn} |n\rangle_f ,
\]  
one obtains  
\[
\hat\rho_i=\frac{1}{Z_i}\sum_{m,n,n'} e^{-\beta_i E_m}\, U_{mn} U_{mn'}^*\, |n\rangle_f \langle n'|_f .
\]  

The quench spectral function can then be expressed as  
\[
S(k,\omega)=\frac{(2\pi)^{2D+1}}{L^D} 
\sum_{n,n',m} \rho_i^{\,n'n}
\langle n|\hat O_1^\dagger|m\rangle 
\langle m|\hat O_2|n'\rangle
\delta(E_n-E_{n'}-\hbar\omega)\, ,
\]  
with matrix elements $\rho_i^{\,n'n} = \frac{1}{Z_i}\sum_m e^{-\beta_i E_m} U_{mn}U_{mn'}^*$.  

The contributions to the off-diagonal density-matrix elements related to the excitation spectral siganl are  
\[
\rho_i^{\,n0} \simeq \frac{1}{Z_i}\sum_m e^{-\beta_i E_m} U_{mn} U_{m0}^* .
\]  
In the weak-quench limit, where $|U_{mm}-1|\ll 1$ and $|U_{mn}|\ll 1$ for $m\neq n$,the leading terms correspond to $m=n$ or $m=0$, giving  
\[
\rho_i^{\,n0} \approx \frac{1}{Z_i}\left( e^{-\beta_i E_0}U_{0n} U_{00}^* + e^{-\beta_i E_n} U_{nn}U_{n0}^*\right).
\]  

Compared with the zero-temperature case (where only the $m=0$ term survives), the finite-temperature initial state introduces an additional term weighted by Boltzmann factor. This modifies only the \textbf{intensity} of the spectral peak, while the excitation energies $E_n$ appearing in the $\delta$-function remain unchanged.  

Thus, finite temperature leads to a redistribution of spectral weight but does not alter the positions of the spectral features themselves. 

In our experiment, finite-temperature effects are indeed present. Based on previous calibrations performed under the same conditions, we estimate the temperature of the system to be about $30\,\mathrm{nK}$~\cite{qihuang2025}, while the recoil energy corresponds to $E_r/k_B \simeq 90\,\mathrm{nK}$ and the initial Hamiltonian for Mott regime has a $U/k_B \simeq0.5E_r/k_B=45\mathrm{nK}$. Thus, the initial state is more accurately described as a thermal state rather than an ideal zero-temperature ground state, however this doesn't affect the spectrum information as discussed above.

We note that a more quantitative analysis would require numerical simulations of thermal states, e.g.\ using purification-based TEBD methods. While this approach is in principle feasible, it is computationally demanding and may suffer from numerical instabilities, which is why we have not included such results here.

\section{The effect of atom number distributions over different tubes}
\label{SM sec.5}

Before loaded into optical lattice, the BEC we prepared has an atom number about $N=8.0 \times 10^4$. The frequency of the optical trap is $\omega_x=\omega_y=2\pi\times24~\rm{Hz}$ and $\omega_z=2\pi\times34~\rm{Hz}$. Then we can have the chemical potential at the center of the BEC:
\[
\mu=\frac{\hbar\omega_{ho}}{2}\left(\frac{15Na}{a_{ho}}\right)^{2/5}
\]
with $\omega_{ho}=\left(\omega_x\omega_y\omega_z\right)^{1/3}$ and $a_{ho}=\sqrt{\hbar/m\omega_{ho}}$. The density distribution can be written as:
\[
n_{3D}\left(x,y,z\right)=\frac{\mu-V\left(x,y,z\right)}{\tilde{g}_{3D}}
\]
in which $\tilde{g}_{3D}=C_lg_{3D}$ represents the rescaled coupling constant~\cite{qihuang2025}. $V(x,y,z)$ in above equation can be written as:
\[
V\left(x,y,z\right)=\frac{1}{2}m\omega_x^2x^2+\frac{1}{2}m\omega_y^2y^2+\frac{1}{2}m\omega_z^2z^2
\]
Integrating along $x$ direction, we can get the planar density distribution:
\[
n_{2D}\left(y,z\right)=\int_xn_{3D}\left(x,y,z\right){\rm d}x=\frac{2m\omega_x^2}{3\tilde{g}_{3D}}\left(R_x^2-y^2-2z^2\right)^{\frac{3}{2}}
\]
with $R_x=\sqrt{2\mu/m\omega_x^2}$. 

Then we load the BEC into optical lattice with lattice depth $V_y=V_z=40\Er$ and lattice period $a=532~\rm{nm}$. Assuming that after the loading of lattices, the BEC is keeping its initial distribution $n_{2D}$ and divided into $m\times n$ sites with area $a\times a$, then the atom number of in site $(i,j)$ can be derived as:
\[
N\left(i,j\right)=\int_{y_i}^{y_{i+1}}\int_{z_j}^{z_{j+1}}n_{2D}\left(y,z\right){\rm d}y~{\rm d}z
\]
with $y_i$ and $z_i$ the coordinates of the site $(i,j)$ and $y_{i+1}-y_i=z_{j+1}-z_j=a$.

By calculating the atom number in each tube, we can using the atom number as weight to evaluate the average atom number of the 1D gas, which can be calculated as:
\[
\bar{N}=\displaystyle \sum_{i,j}\frac{N\left(i,j\right)^2}{N}
\]
In our experiment, $\bar{N}=45$.

Using the above distribution, we sample the different atom numbers of tubes along the long axis of the area to do DMRG simulation to get the Quench Spectral Function, and average them with the weight of distances from the tubes to the center tube. Specifically, we carry out simulations for two representative experimental cases: a quench from $V_z = 24.5E_r$ to $V_z = 22E_r$ in the Mott regime (subplots (a1), (a2)), and a quench from $V_z = 6E_r$ to $V_z = 5E_r$ in the superfluid regime (subplots (b1), (b2)).

\begin{figure}[t]
    \centering
    \includegraphics[width=0.7\linewidth]{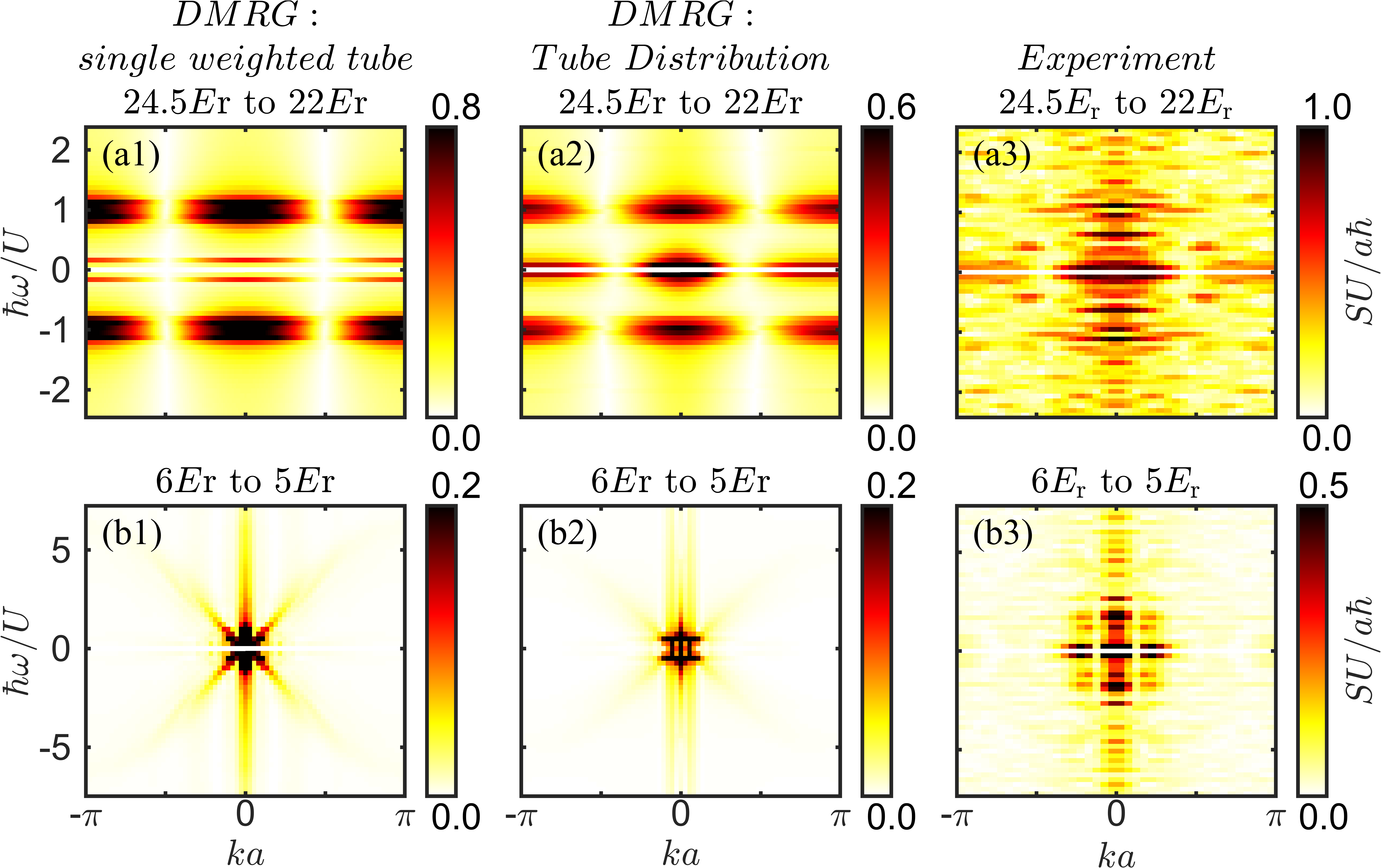}
    \caption{Amplitudes of quench spectral functions for $V_z=24.5E_r\rightarrow22E_r$ (Mott) quenches (a1)-(a3) and $V_z=6E_r\rightarrow5E_r$ (superfluid) quenches (b1)-(b3). (a1)(b1) single-tube DMRG simulation using effective weighted atom number; (a2)(b2) averaged multi-tube DMRG simulation using distribution of atom numbers; (a3)(b3) experimental results.}
    \label{fig:replyFig1}
\end{figure}

As shown in Fig.~\ref{fig:replyFig1}, in the Mott-regime case we find that an elliptical shape near zero frequency can arise from the sum over of tubes with different atom numbers. This fits better with the measured experimental data (a3) which confirms our initial interpretation "Notably, the signals at $\hbar \omega = 0$ is also broadened into an ellipse shape, which is probably due to the tube distributions in actual experiment". Importantly, the main spectral features and excitation energies remain essentially unaffected. Also, by comparing Fig.~\ref{fig:replyFig1} (b1) and (b2), a similar robustness is observed in the superfluid case.

\section{Analysis of Experimental data}
\label{SM sec.6}
In this section, we complement some details about error analysis of experimental data, analysis of Mott gap and the cutoff at $ka=\pi/2$ on our experimental data. 

\vspace{1\baselineskip}

\textit{Error analysis of experimental data.}

\begin{figure}[tb]
    \centering
    \includegraphics[width=0.8\linewidth]{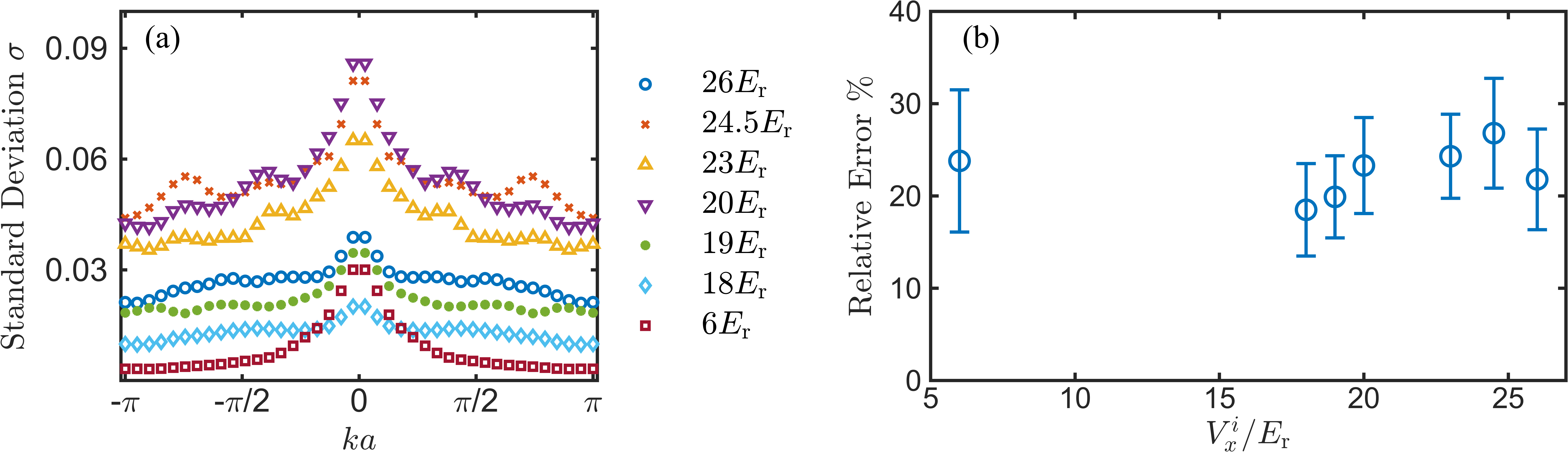}
    \caption{(a) Standard deviation of experimental data. Each line corresponds to a quench parameter. The data point at momentum $k$ represents the standard deviation of the $k$ column of $S(k,\omega)$; (b) Average relative error of $S(k,\omega)$. The data point in $S(k,\omega)$ lower than $10\%$ of maximum is excluded in the calculation.}
    \label{fig:replyFig_STD}
\end{figure}

 In our experiment, we obtain $6$ TOF pictures at each evolution time step. Then we can get evolution of momentum distribution $n(k,t)$ and the standard deviation of each data point $E(k,t)$. After that, $S(k,\omega)$ is obtained by performing Discrete Fourier Transform(DFT) to each column of $n(k,t)$ and the standard deviation of each column can be calculated according to $E(k,t)$. Fig.\ref{fig:replyFig_STD}(a) shows the standard deviation of $S(k,\omega)$ from the averaging of these 6 groups of data. Each line in this figure corresponds to a quenching parameter we measured. The data point at position $k$ represents the standard deviation of the corresponding column in $S(k,\omega)$. According to these results and $S(k,\omega)$, the relative error of $S(k,\omega)$ can be calculated and the result is shown in Fig.\ref{fig:replyFig_STD}(b). Each point in this figure represents the average relative error of the corresponding quenching parameter. In this calculation, the data point in $S(k,\omega)$ lower than $10\%$ of maximum is excluded.  We can see that the average relative error of $S(k,\omega)$ is around $20 \%$ .

\vspace{1\baselineskip}

\textit{Analysis of Mott gap.}

\begin{figure}[t]
    \centering
    \includegraphics[width=1.0\linewidth]{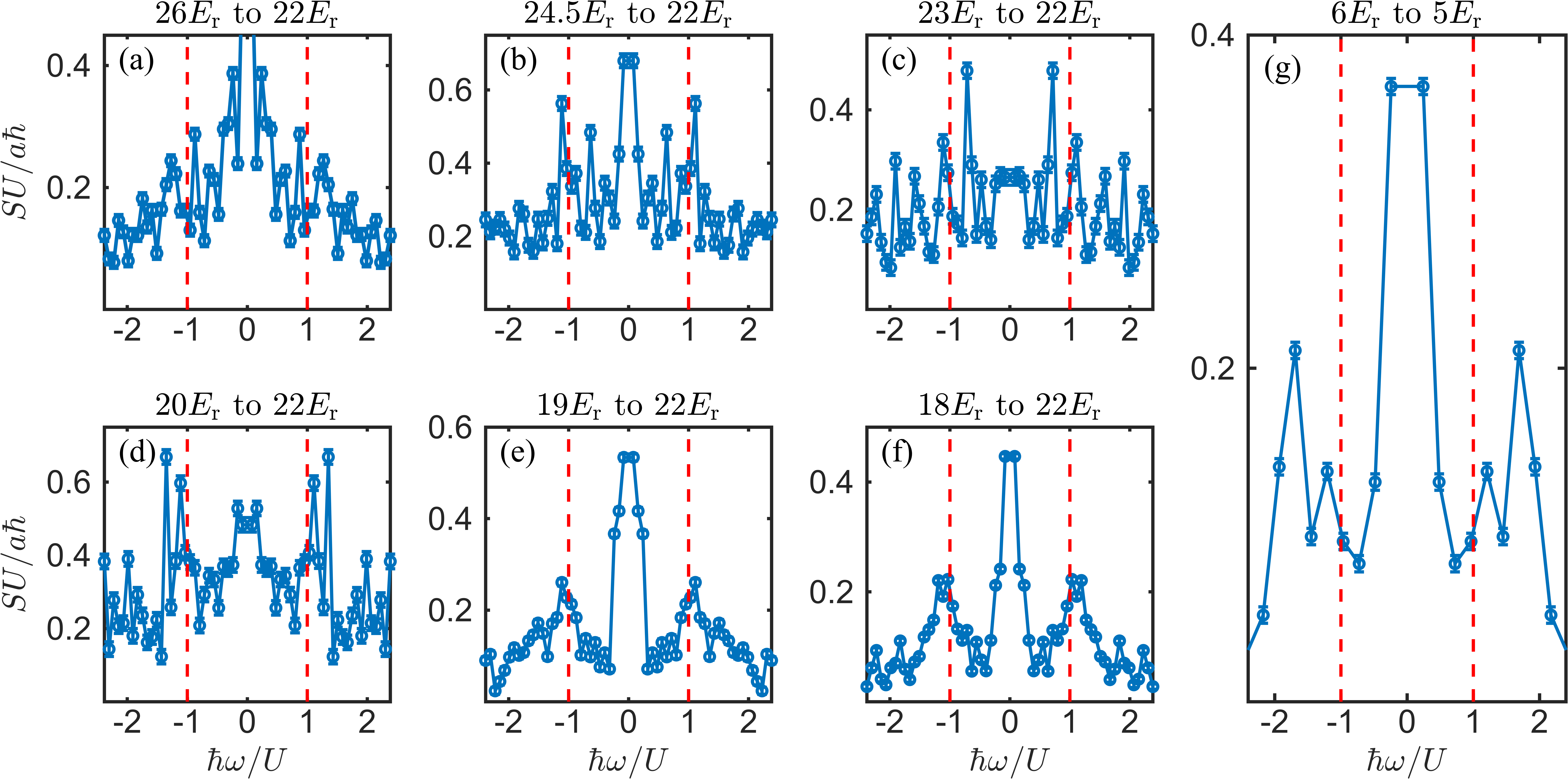}
    \caption{Integral of $S(k,\omega)$ between $ka=0$ and $ka=\pi/2$. (a)-(c) correspond to quenches from deep lattice $26\Er$ to $22\Er$, $24.5\Er$ to $22\Er$ and $23\Er$ to $22\Er$; (d)-(f) represent the result of reverse quench $20\Er$ to $22\Er$, $19\Er$ to $22\Er$ and $18\Er$ to $22\Er$; (g) shows the quench in Super Fluid regime $6\Er$ to $5\Er$. The red dash line show the position of $\hbar\omega/U=\pm1$.}
    \label{fig:replyFig_gapA}
\end{figure}

To quantitatively analysis the Mott gap, we calculate the integral of $S(k,\omega)$ between $k=0\sim k_e$:
\[
    S_\omega(\omega)=\frac{1}{k_e}\int_0^{k_e}S(k,\omega){\rm d}k
\]
where $k_e$ represents the maximum $k$ position the energy spectrum extend to. Here, we set $k_ea=\pi/2$. Fig.\ref{fig:replyFig_gapA} shows the calculated results for different quench parameters. The red dash line indicate the position $\hbar\omega/U=1$. For (a)-(c), We can see two peaks near $\hbar\omega/U=1$, while for (d)-(f) we can also observe a broad peak at $\hbar\omega/U=1$. As a comparison, (g) shows the result of the quench sperctrum of superfluid phase. There is no obvious peak at $\hbar\omega/U=1$. On the contrary, a signal at $\hbar\omega>U$ is observed and it fits with the Bogoliubov excitations of superfluid. According to these results, we conclude that a Mott gap around $\hbar\omega=U$ is observed for the QSF in Mott insulator regimes. 
    
\begin{figure}[t]
    \centering
    \includegraphics[width=1.0\linewidth]{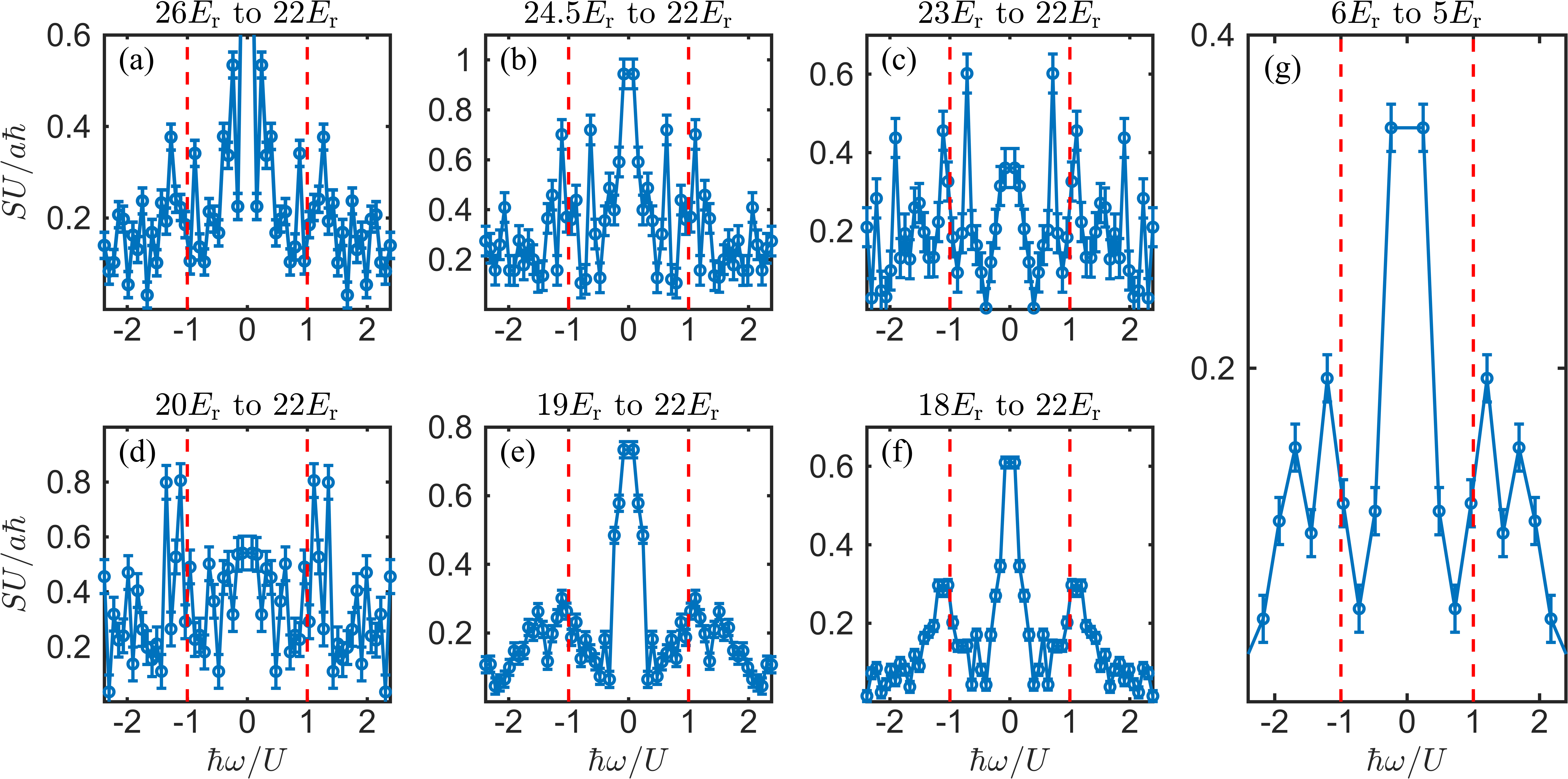}
    \caption{Cut of $S(k,\omega)$  at $ka=\pi/5$. (a)-(c) correspond to quenches from deep lattice $26\Er$ to $22\Er$, $24.5\Er$ to $22\Er$ and $23\Er$ to $22\Er$; (d)-(f) represent the result of reverse quench $20\Er$ to $22\Er$, $19\Er$ to $22\Er$ and $18\Er$ to $22\Er$; (g) shows the quench in Super Fluid regime $6\Er$ to $5\Er$. The red dash line show the position of $\hbar\omega/U=\pm1$.}
    \label{fig:replyFig_gapcutA}
\end{figure}

Here we also give the cut result we previously used to analysis the Mott gap. Fig.\ref{fig:replyFig_gapcutA} shows the cut of $S(k,\omega)$ at $ka=\pi/5$ for different quench parameters. From this figure we can see that, although the height of the peaks are different from the integral results, the existence and position of the peaks are almost the same as the corresponding integral results shown in Fig.\ref{fig:replyFig_gapA}.

\begin{figure}[htbp]
    \centering
    \includegraphics[width=0.8\linewidth]{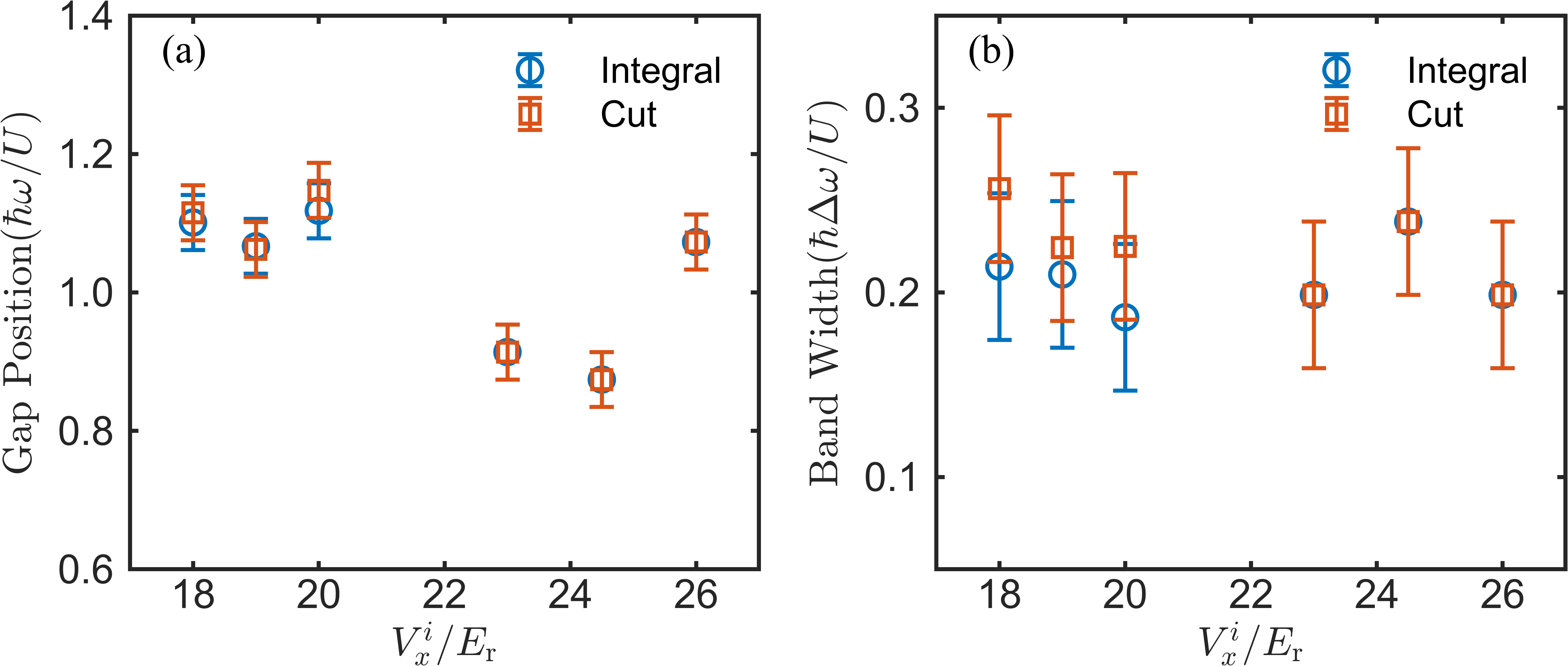}
    \caption{Gap position and band width extracted from Fig.\ref{fig:replyFig_gapA} (the blue circle) and Fig.\ref{fig:replyFig_gapcutA} (the orange square). (a) Gap position by calculating $(\omega_{ub}+\omega_{lb})/2$; (b) Band width by calculating $(\omega_{ub}-\omega_{lb})/2$. Here $\omega_{ub}$ and $\omega_{lb}$ represent the upper bound and lower bound of the Mott gap, respectively. The error bar is evaluated from frequency resolution of $S(k,\omega)$.}
    \label{fig:replyFig_widthA}
\end{figure}

From the integral and cut data shown above, the upper bound $\omega_{ub}$ and lower bound $\omega_{lb}$ of the Mott gap can also be extracted. For subplots (a)-(c), we can use the two peaks around $\hbar\omega/U=1$ as the bound of Mott gap. For subplots (d)-(f), the FWHM can also be a reasonable estimator of the bound. Then we can calculate $(\omega_{ub}+\omega_{lb})/2$ to evaluate the position of the Mott gap. Fig.\ref{fig:replyFig_widthA}(a) shows the gap position of Mott quenches. The blue circle is read from integral of $S(k,\omega)$ from $ka=0$ to $ka=\pi/2$. The orange square is read from cut of $S(k,\omega)$ at $ka=\pi/5$. The errors are mainly come from the detection resolution which is decided by the maximum evolution time. In our experiment, it is $t=12~{\rm ms}$ which leads to an uncertainty about $\hbar\sigma_{\omega}/U=\pm0.040$. According to these results, the average position of Mott gap is $\hbar\omega/U=1.024$ with a standard deviation $0.104$ for integral method. And for the cut method, it is $\hbar\omega/U=1.031$ with a standard deviation $0.111$.
    
Using the extracted upper bound $\omega_{ub}$ and lower bound $\omega_{lb}$  of the Mott gap, we can calculate $(\omega_{ub}-\omega_{lb})/2$ to evaluate the width of the band.
Fig.\ref{fig:replyFig_widthA}(b) shows the extracted band width of each Mott quenches. The blue circle is read from integral of $S(k,\omega)$ from $ka=0$ to $ka=\pi/2$. The orange square is read from cut of $S(k,\omega)$ at $ka=\pi/5$. According to the data, for integral result(blue), the average band width of quenches from deep lattice is $\hbar\Delta\omega/U=0.212\pm0.023$ and for reverse quench, the average band width is $\hbar\Delta\omega/U=0.203\pm0.015$. And for cut result(orange), the average band width are $\hbar\Delta\omega/U=0.212\pm0.023$ and $\hbar\Delta\omega/U=0.235\pm0.018$ for deep lattice quench and reverse quench, correspondingly. As a comparison, the band width is $\hbar\Delta\omega/U=0.170$ according to the explanation we proposed. The precision of the band width we calculate is mainly limited by the detection resolution which is decided by the maximum evolution time. In our experiment, it is $t=12~{\rm ms}$, which leads to an uncertainty about $\hbar\sigma_{\omega}/U=\pm0.040$.

\vspace{1\baselineskip}

\textit{Analysis of the cutoff at $ka=\pi/2$.}

\begin{figure}[t]
    \centering
    \includegraphics[width=0.8\linewidth]{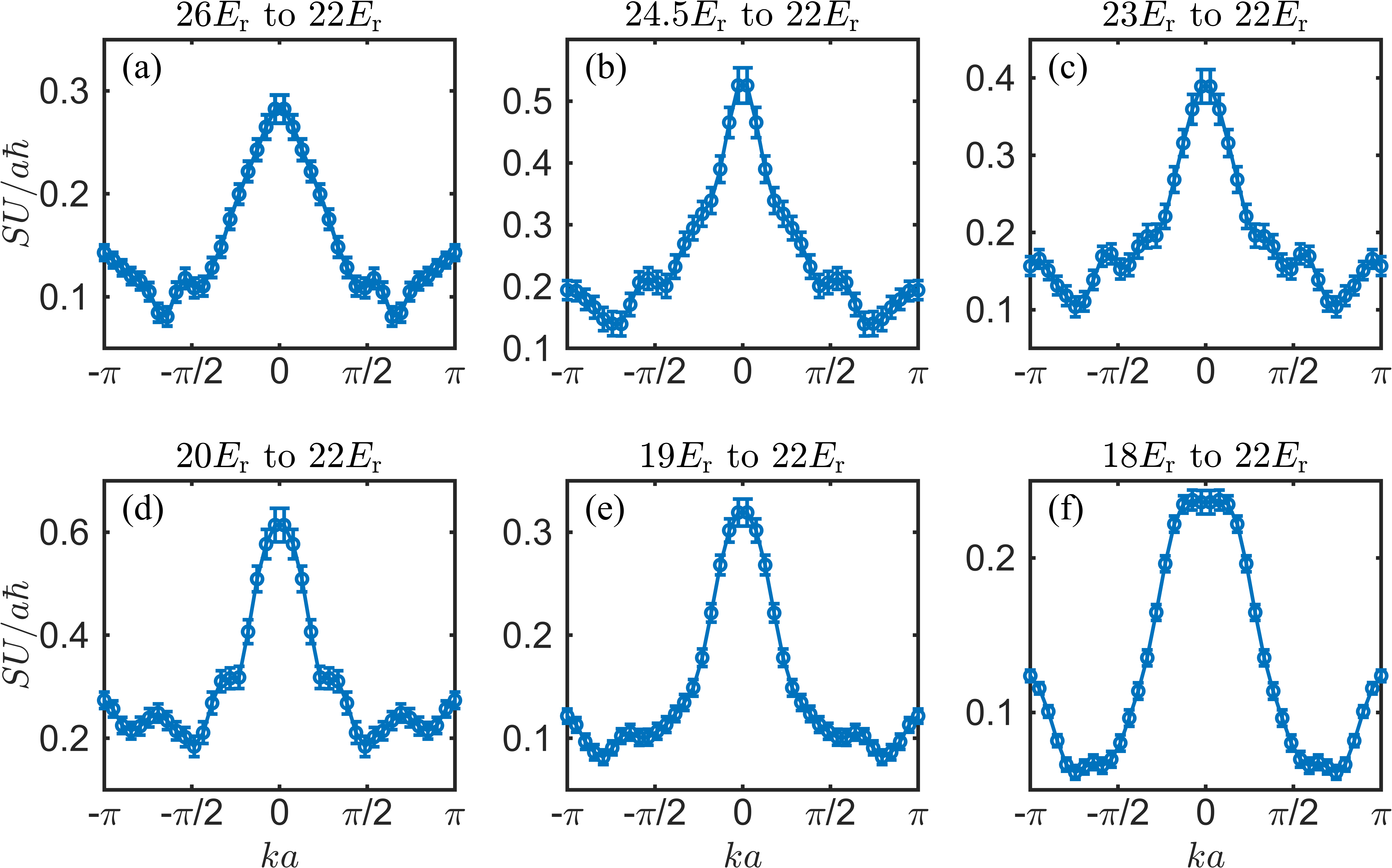}
    \caption{Integral of $S(k,\omega)$ between $\hbar\omega_d/U=0.7$ and $\hbar\omega_u/U=1.3$. (a)-(c) correspond to quenches from deep lattice $26\Er$ to $22\Er$, $24.5\Er$ to $22\Er$ and $23\Er$ to $22\Er$; (d)-(f) represent the result of reverse quench $20\Er$ to $22\Er$, $19\Er$ to $22\Er$ and $18\Er$ to $22\Er$.}
    \label{fig:replyFig_cutoff}
\end{figure}

\begin{figure}[b]
    \centering
    \includegraphics[width=0.5\linewidth]{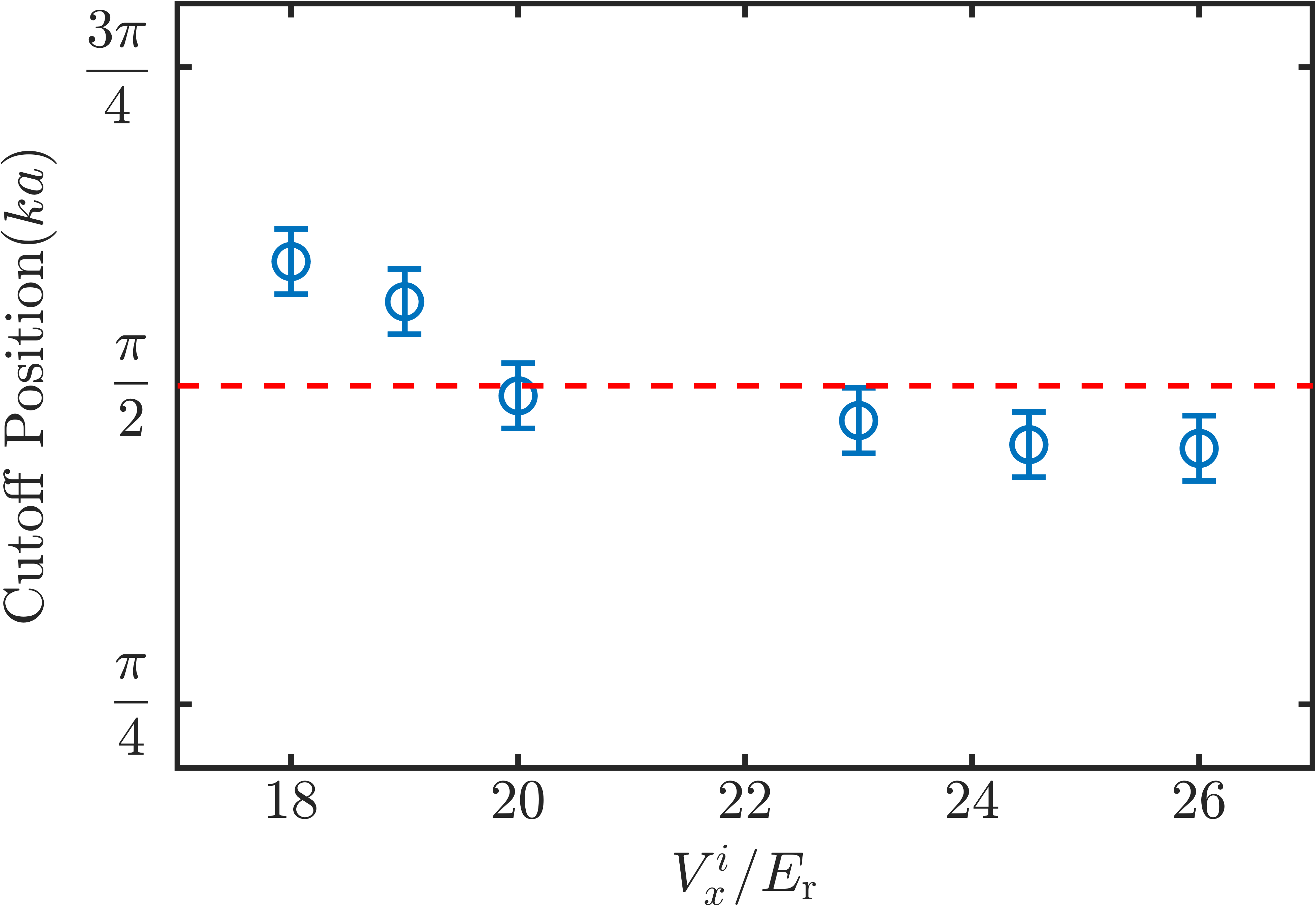}
    \caption{Cutoff position extracted from Fig.\ref{fig:replyFig_cutoff}. It is calculated by finding the zero point of $S_k^{(1)}(k)$ closest to $ka=0$. The error bar is evaluated from the momentum resolution of $S(k,\omega)$. The red dash line indicates the position $ka=\pi/2$.}
    \label{fig:replyFig_CutoffPos}
\end{figure}

To quantitatively observe the cutoff around $ka=\pi/2$, similarly we can calculate the integral of $S(k,\omega)$ between $\omega=\omega_d$ to $\omega=\omega_u$:
\[
    S_k(k)=\frac{1}{\omega_u-\omega_d}\int_{\omega_d}^{\omega_u}S(k,\omega){\rm d}\omega
\]
in which the integral range $\omega_d\sim \omega_u$ contain the Mott gap. Fig.\ref{fig:replyFig_cutoff} shows the results of Mott quenches. Here we choose $\hbar\omega_d/U=0.7$ and $\hbar\omega_u/U=1.3$, which is wide enough to contain the Mott gap according to DMRG results. From this figure, we can see that for all quench parameter in our experiment, the energy spectrum have a cutoff around $ka=\pi/2$. To extract the cutoff position, we calculate the first-order difference of $S_k(k)$:
\[
    S_k^{(1)}(k_i)=S_k(k_{i+1})-S_k(k_i)
\]
Then we calculate the position of the zero point which is closest to $ka=0$. The result of each quenching parameter is shown in Fig.\ref{fig:replyFig_CutoffPos}. The error of the data points mainly come from momentum resolution. In our experiment, it is about $\sigma_{k}a= \pm0.081\approx\pm5\%\times\pi/2$. This results allow us to say that there exists a cutoff around $ka=\pi/2$ in the energy spectrum.

\end{document}